\documentclass[useAMS,usenatbib,letterpaper]{mn2e}
\usepackage{graphicx}
\usepackage{txfonts}
\usepackage[authoryear]{natbib}
\bibpunct[,]{(}{)}{;}{a}{,}{,}

\begin{document}
\title[A grid of Chemical evolution models]{A grid of 
chemical evolution models as a tool to interpret spiral 
and irregular galaxies data. }

\author[Moll\'{a} \&
D\'{\i}az]{M. Moll\'{a}$^{1}$\thanks{E-mail:mercedes.molla@ciemat.es}
\and A. I. D\'{\i}az$^{2}$\thanks{E-mail:angeles.diaz@uam.es}\\
$^{1}$Departamento de Fusi\'{o}n y F\'{\i}sica de 
Part\'{\i}culas Elementales. C.I.E.M.A.T Avda. Complutense 22, 
28040, Madrid, (Spain) \\
$^{2}$Departamento de F\'{\i}sica Te\'{o}rica,
Universidad Aut\'onoma de Madrid, 28049 Cantoblanco, Madrid (Spain)}

\date{Accepted Received ; in original form }

\pagerange{\pageref{firstpage}--\pageref{lastpage}} \pubyear{2004}

\maketitle
\label{firstpage}

\begin{abstract}

We present a generalization of the multiphase chemical evolution model
applied to a wide set of theoretical galaxies with different masses
and evolutionary rates.  This generalized set of models has been
computed using the so-called {\sl Universal Rotation Curve} from
\citet*{per96} to calculate the radial mass distribution of 44 {\sl
theoretical} protogalaxies. This distribution is a fundamental input
which, besides its own effect on the galaxy evolution, defines the
characteristic collapse time-scale or gas infall rate onto the disc.
We have adopted 10 sets of values, between 0 and 1, for the
molecular cloud and star formation efficiencies, as corresponding to
their probability nature, for each one of the radial distributions of
total mass. Thus, we have constructed a bi-parametric grid of models,
depending on those efficiency sets and on the rotation velocity,
whose results are valid in principle for any spiral or irregular
galaxy.  The model results provide the time evolution of different
regions of the disc and the halo along galactocentric distance,
measured by the gas (atomic and molecular) and stellar masses, the
star formation rate and chemical abundances of 14 elements, for a
total of 440 models.  This grid may be used to estimate the evolution
of a given galaxy for which only present time information -- such as
radial distributions of elemental abundances, gas densities and/or
star formation, which are the usual observational constraints of
chemical evolution models -- is available.

\end{abstract}


\section{Introduction}

Chemical evolution models (CEM) \citep{lyn75,tin80,cla87,cla88,som89}
were early developed to try to understand the origin of the radial
gradients of abundances, observed in our Galaxy (MWG).  Most numerical
models in the literature, including the multiphase model used in this
work, explain the existence of this radial gradient by the combined
effects of a star formation rate (SFR) and an infall of gas which vary
with galactocentric radius in the Galaxy.

A radial decrease of abundances has also been observed in most
spiral galaxies \citep{hen99} although the shape of the radial
distribution changes from galaxy to galaxy. Among other global trends
it is found that for isolated non-barred spirals the steepness of the
radial gradient depends on morphological type, with later types
showing steeper gradients \citep{diaz89}, with other general galaxy
properties as surface brightness and neutral and molecular gas
fractions also playing a role \citep{vil92,zar94}. The radial gradient
tends to be wiped out however for strongly barred galaxies which show
flat abundance distributions \citep{mar94,roy96,fri94}. Irregulars
galaxies also show uniform abundances throughout
\citep{roy96b,walsh97,kob98,mol99b}.

The abundance gradient pattern seems to show an on-off mode
\citep{edm93}, being very steep for the latest spiral types and very
flat for irregulars. All these considerations become clear when the
gradient is measured in dex/kpc, but there are indications that
suggest a gradient independent of galaxy type when it is measured in
dex/scale length \citep{diaz89,gar98}. In order to analyze the
behaviour of the radial distribution of abundances and the value of
the radial gradient from a theoretical point of view a large number of
models is necessary.  Historically, CEM aiming to reproduce radial
abundance gradients have been, however, applied only to the MWG.

Actually, there is a lack of tools to determine the chemical
evolutionary state of a particular galaxy, besides our works
applying the multiphase models to spiral galaxies.  The recent works by
\citet{boi00,boi00b} are valid for galaxies other than the MWG. Their
calculations use the angular momentum and rotation curves as model
inputs keeping the star formation efficiency constant for all galaxies
\citep*{boi01}. This technique may not be flexible enough to validate
the models against observational data. In fact, a comparison to see if
these models reproduce the observed abundance radial distributions of
particular galaxies has not been done.  It is always possible to
extract some information by using evolutionary synthesis models in
comparison with spectro-photometric observations. This method, very
useful for the study of elliptical galaxies, does not result equally
successful in the case of spiral galaxies due to the difficulty of
measuring the spectral indices, except for the bulges
\citep{gou99,proc02}, from which ages and metallicities are obtained.
Furthermore, even when these measurements are done with confidence
\citep{beau97,mol99}, in order to apply this technique to spiral
galaxies, a combination of chemical evolution and evolutionary
synthesis models is required to solve the uniqueness problem associated
to the first ones and the age-metallicity degeneracy associated to the
second ones \citep{mol02}.

At present, the available options are either to use the classical
closed box model or a Galactic Chemical Evolution (GCE)
model. However, the closed box scenario is recognised to be inadequate
to describe the evolution of most galaxies and in fact its application
in many cases can yield misleading results \citep*{vall02}.  In
particular, the fact of assuming that a system has a constant total
mass with a monotonically decreasing star formation according to a
Schmidt law, prevents the reproduction of the observational
characteristics of most galaxies.  On the other hand, the evolution of
a galaxy with present time properties different from the Milky Way
will not necessarily be equal to that predicted by a GCE model.
Realistic chemical evolution models adequate to describe different
types of spiral and irregular galaxies are therefore clearly needed.

The multiphase model, whose characteristics have been described in
\citet*{fer92}, has been applied and checked against observational
constraints not only for the Milky Way Galaxy \citep*{fer94,mol95}, as
it is commonly done, but also for a sample of spiral galaxies (discs
and bulges) of different morphological types and total masses
\citep*{mol96,mol99,mol99b,mol00}. The observed radial distributions
of gas, oxygen abundances and star formation rate have been reproduced
rather successfully and the observed correlations between abundance
gradients and galaxy characteristics are also reproduced
\citep{mol96,mol99c}. This galaxy sample, which includes the best
studied objects, is however small (only 11) and encompasses a
restricted range of morphologies and masses. The application of the
model can however be extended to a larger sample if an adequate
parameter space is defined thus providing the required chemical
evolution of different types of galaxies.

The model uses as input parameters the collapse time scale to form the
disc, which depends on the total mass of the galaxy, and the
efficiencies to form molecular clouds and stars which we assume
different from galaxy to galaxy.  The radial distributions of total
mass constitute the fundamental input of the multiphase model.  They
are easily computed when the rotation curves are available (Moll\'{a}
\& M\'{a}rquez, in preparation).  If this is not the case, some
assumptions are necessary. In this work, we have used the Universal
Rotation Curve from \citet[][hereafter PSS96]{per96} to calculate a
large number of mass radial distributions representing theoretical
protogalaxies or initial structures which will evolve to form the
observed spiral discs or irregulars.  The total mass of each simulated
galaxy, besides having its own effect on the galaxy evolution, defines
the characteristic collapse time-scale or gas infall rate onto the
disc.  Regarding molecular cloud and star formation efficiencies,
which will take values between 0 and 1, we have chosen 10
different sets of values for each radial distribution of total mass.
We have computed how the chemical evolution proceeds for galaxies
defined by the different parameter combinations.  The final
bi-parametric grid consists of 440 models simulating galaxies of 44
different total masses.

This work represents an extension of our previous work that can help
to understand the general trends observed in spiral and irregular
galaxies concerning neutral and molecular gas distributions, abundance
radial distributions etc. But, most importantly, by using these
models we can predict the time evolution of a galaxy when only present
time data are known.

In Section 2 we summarize the general characteristics of the
multiphase chemical evolution model and the strategy of its
application to the objects of the grid.  The results are presented in
Section 3 including the time evolution of galaxies and some
radial distributions for the present time. In section 4
we give the calibration of the grid models with the MWG
and a restricted sample of well studied spiral galaxies 
and we discuss the model results in a global way.  Finally, the
conclusions of this work are presented in Section 5.

The results obtained in this work will be available in electronic form
at CDS via anonymous ftp to cdsarc.u-strasbg.fr (130.79.128.5), via
{http://cdsweb.u-strasbg.fr/Abstract.html}, or at
{http://wwwae.ciemat.es/$^{\sim}$mercedes}.

\section{The Multiphase Model}

\begin{figure}
\resizebox{\hsize}{!}{\includegraphics[angle=0]{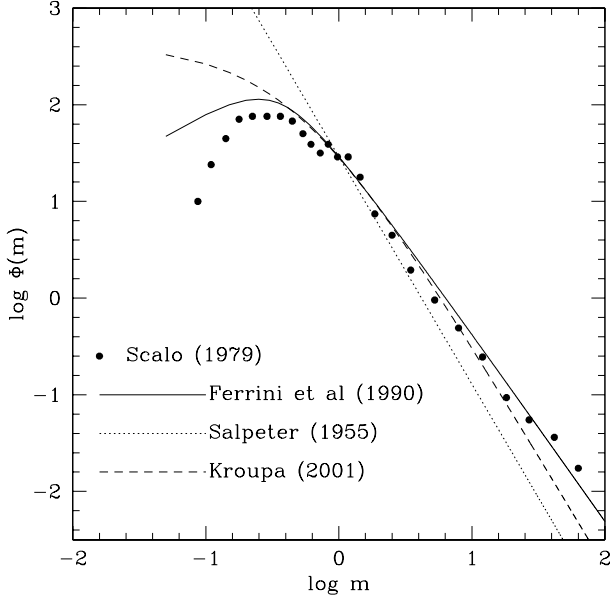}}
\caption{The IMF from \citet{fer90}, solid line, compared to a
Salpeter law, dotted line, and that corresponding to \citet{kro01}, 
short-dashed line. Solid symbols correspond to \citet{sca86}.}
\label{imf}
\end{figure}

The model used in this work is a generalization of that developed for
the Solar Neighborhood in \citet{fer92} and later applied to the whole
MWG \citep{fer94} and other spiral galaxies \citep{mol96,mol99}.  The
enriched material proceeds from the restitution by dying stars,
considering their nucleosynthesis, their initial mass function (IMF)
-- and hence the delayed restitution -- and their final fate, via a
quiet evolution, or Type I and II supernova explosions.  Most recent
works support the idea that the IMF is practically universal in space
and constant in time \citep*{wys97,sca98,mey00}, showing only local
differences.  We have adopted the IMF from \citet{fer90}, very similar
to a Scalo's law \citep{sca86} and in good agreement with the most
recent data from \citet{kro01}, as can be seen in Fig.~\ref{imf}.

The original model has been modified in order to use metallicity
dependent yields. Nucleosynthesis yields for massive stars have been
taken from \citet{woo95}. For low mass and intermediate stars we have
used the set of yields from \cite*{gav04}.  For type I supernova
explosion releases, the model W7 from \cite{nom84}, as revised by
\citet{iwa99} has been taken.

\subsection{The rotation curves and mass radial distributions}

In this model each galaxy is described as a two-zone system with a
halo and a disk. It is assumed that the halo has a total mass which is
initially in gas phase.  The total mass, M, and its radial
distribution, M(R), are calculated from the corresponding rotation
curve derived from the Universal Rotation Curve of PSS96.  These
authors use a homogeneous sample of about 1100 optical and radio
rotation curves to estimate their profile and amplitude which are
analysed statistically. From this study they obtain
an expression for the rotation velocity, V(R), as a function of the
rotation velocity at the optical radius (the radius encompassing 83\%
of the total integrated light), $V_{\mathrm{opt}}$, the galaxy radius
normalised to the optical one, $x=R/R_{\rm opt}$, and a parameter
$\lambda = L/L_\mathrm{*}$, which represents the ratio of the galaxy
luminosity to that of the MWG, 
$L_\mathrm{*} = 10^{10.4} L_\mathrm{\odot}$:

\begin{equation}
V(R)=V_{\mathrm{opt}} F(\lambda) \quad \mathrm{km s^{-1}}
\label{crot}
\end{equation}

where
\begin{equation}
F(\lambda)=\left\{ (0.72+0.44 \log{\lambda}) \frac{1.97
x^{1.22}}{(x^{2}+0.61)^{1.43}} + 1.6 e^{-0.4 \lambda}
 \frac{x^{2}}{x^{2}+2.25 \lambda^{0.4}}\right\}^{1/2} 
\end{equation}

These authors also found that $V_{\mathrm{opt}}$
in $\rm{ km s^{-1}}$, depends on $\lambda$ as :

\begin{equation}
V_{\mathrm{opt}}=\frac{200 \lambda^{0.41}}{[0.80 + 0.49 log \lambda +
 (0.75 e^{-0.4 \lambda})/(0.47 + 2.25 \lambda^{0.4})]^{1/2}} 
\label{vopt}
\end{equation} 

The same occurs with  the optical radius, which depends on
luminosity through the expression:  
$R_{\rm opt} = 13 (L/L_{\rm *})^{0.5}$ kpc, which we will also use.

\begin{figure}
\resizebox{\hsize}{!}{\includegraphics[angle=0]{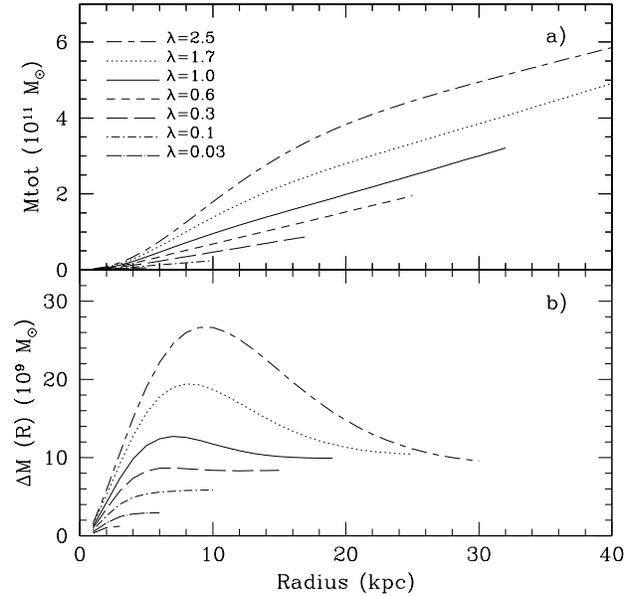}}
\caption{Radial distributions: a) total masses M$_{tot}$, b) masses
$\Delta M (R)$ included in our cylinders, for different values of
$\lambda$ following the labels in the figure.}
\label{dis}
\end{figure}

\begin{table}
\caption{Galaxy Characteristics dependent on the Total Mass.}
\label{grid_m}
\begin{tabular}{rlcccccc}
\hline
N & $\lambda$ & $R_{opt}$ & R$_{\rm gal}$ & 
R$_{c}$ & V$_{\rm max}$ & M$_{gal}$ & $\tau_{c}$ \\  
    &  & (kpc) & (kpc) & (kpc) & ($\rm km s^{-1}$) &
 (10$\rm ^{9}M_{\odot})$& (Gyr)\\
\noalign{\smallskip}
\hline
\noalign{\smallskip}
 1& 0.01 &  1.3&  36.7&  0.7&  30.&     8.& 60.37\\ 
 2& 0.02 &  1.8&  47.1&  0.9&  40.&    18.& 40.12\\ 
 3& 0.03 &  2.3&  54.4&  1.1&  48.&    29.& 31.59\\ 
 4& 0.04 &  2.6&  60.4&  1.3&  54.&    40.& 26.66\\ 
 5& 0.05 &  2.9&  65.4&  1.5&  59.&    52.& 23.38\\ 
 6& 0.06 &  3.2&  69.9&  1.6&  63.&    65.& 21.00\\ 
 7& 0.07 &  3.4&  73.9&  1.7&  67.&    78.& 19.17\\ 
 8& 0.08 &  3.7&  77.5&  1.8&  71.&    91.& 17.72\\ 
 9& 0.09 &  3.9&  80.9&  1.9&  75.&   105.& 16.53\\ 
10& 0.10 &  4.1&  84.0&  2.1&  78.&   119.& 15.54\\ 
11& 0.11 &  4.3&  86.9&  2.2&  81.&   133.& 14.69\\ 
12& 0.12 &  4.5&  89.7&  2.3&  84.&   147.& 13.96\\ 
13& 0.13 &  4.7&  92.3&  2.3&  87.&   162.& 13.31\\ 
14& 0.14 &  4.9&  94.8&  2.4&  90.&   176.& 12.74\\ 
15& 0.15 &  5.0&  97.2&  2.5&  92.&   191.& 12.24\\ 
16& 0.16 &  5.2&  99.5&  2.6&  95.&   207.& 11.78\\ 
17& 0.17 &  5.4& 101.7&  2.7&  97.&   222.& 11.37\\ 
18& 0.18 &  5.5& 103.8&  2.8&  99.&   237.& 10.99\\ 
19& 0.19 &  5.7& 105.8&  2.8& 101.&   253.& 10.64\\ 
20& 0.20 &  5.8& 107.8&  2.9& 104.&   269.& 10.33\\ 
21& 0.30 &  7.1& 124.7&  3.6& 122.&   433.&  8.13\\ 
22& 0.40 &  8.2& 138.3&  4.1& 138.&   608.&  6.86\\ 
23& 0.50 &  9.2& 149.9&  4.6& 151.&   791.&  6.02\\ 
24& 0.60 & 10.1& 160.1&  5.0& 163.&   981.&  5.41\\ 
25& 0.70 & 10.9& 169.2&  5.4& 173.&  1176.&  4.94\\ 
26& 0.80 & 11.6& 177.5&  5.8& 183.&  1377.&  4.56\\ 
27& 0.90 & 12.3& 185.2&  6.2& 192.&  1582.&  4.26\\ 
28& 1.00 & 13.0& 192.4&  6.5& 200.&  1791.&  4.00\\ 
29& 1.10 & 13.6& 199.1&  6.8& 208.&  2004.&  3.78\\ 
30& 1.20 & 14.2& 205.5&  7.1& 216.&  2220.&  3.59\\ 
31& 1.30 & 14.8& 211.5&  7.4& 223.&  2440.&  3.43\\ 
32& 1.40 & 15.4& 217.2&  7.7& 230.&  2663.&  3.28\\ 
33& 1.50 & 15.9& 222.6&  8.0& 236.&  2888.&  3.15\\ 
34& 1.60 & 16.4& 227.9&  8.2& 243.&  3117.&  3.03\\ 
35& 1.70 & 16.9& 232.9&  8.5& 249.&  3347.&  2.93\\ 
36& 1.80 & 17.4& 237.7&  8.7& 255.&  3581.&  2.83\\ 
37& 1.90 & 17.9& 242.4&  9.0& 260.&  3816.&  2.74\\ 
38& 2.00 & 18.4& 246.9&  9.2& 266.&  4054.&  2.66\\ 
39& 2.50 & 20.6& 267.6& 10.3& 291.&  5274.&  2.33\\ 
40& 3.00 & 22.5& 285.7& 11.3& 314.&  6539.&  2.09\\ 
41& 3.50 & 24.3& 302.0& 12.2& 335.&  7843.&  1.91\\ 
42& 4.00 & 26.0& 316.9& 13.0& 353.&  9180.&  1.77\\ 
43& 4.50 & 27.6& 330.6& 13.8& 371.& 10548.&  1.65\\ 
44& 5.00 & 29.1& 343.4& 14.5& 387.& 11944.&  1.55\\ 
\hline
\end{tabular}
\end{table}
\normalsize

In Table~\ref{grid_m} we show the characteristics obtained with these
equations for 44 different values of $\lambda$. In column (1) we give
the number of the radial distribution, defined by the value of
$\lambda$, given in column (2). The optical radius, $\rm R_{opt}$, and
the virial radius, $R_\mathrm{gal}=14.8 \lambda^{-0.14}
R_\mathrm{opt}$, are given in columns (3) and (4) respectively.  
We have defined a characteristic radius for each galaxy, which we will
use as our reference radius, as $R_\mathrm{c}=R_\mathrm{opt}/2$. This
radius, in kpc, is given in column (5). Column (6) gives the rotation
velocity, in $\mathrm {km s^{-1}}$, reached at a radius
$R_\mathrm{M}=R_{opt}/1.45$ kpc (see PSS96 for details).  The total
mass of the galaxy, calculated with the classical expression
$M_\mathrm{gal}=2.32 10^{5} R_\mathrm{gal} V_\mathrm{max}^{2}$
\citep{leq83}, in units of 10$\rm ^{9}$ M$_{\odot}$, is given in
column (7), and, finally, the characteristic collapse time scale, in
Gyr, which will be described below, is given in column (8).

Each galaxy is divided into concentric cylindrical regions 1 kpc wide.
From the corresponding rotation curve we calculate the radial
distributions of total mass, M(R), with the expression $ M(R)=2.32
10^{5} R V(R)^{2}$. \footnote{This equation is valid only for
spherical distributions. The corresponding expression for cylindrical
distributions differs from the spherical one only by a geometrical
factor, between 0 and 1 \citep{bur85,cam93}. This difference is well
inside the possible uncertainties of the actual rotation curves
compared with the universal rotation curve from PSS96.}

From these distributions of the total mass, we easily obtain the one
corresponding to each of the cylinders, $\Delta M(R)$.  Both
distributions $M(R)$ and $\Delta M(R)$ are shown in
Fig.~\ref{dis}. The total mass includes the dark matter component (DM)
which, in principle, does not take part in the chemical
evolution. However,the DM contribution seems to be negligible for the
large massive galaxies, mostly in the regions where the chemical
evolution is calculated. According to \citet[][and references
therein]{pal00,sell00} 75\% of the spiral galaxies are well fitted
without a dark matter halo and the failure to reproduce an other 20\%
is directly related to the existence of non-axisymmetric structures
(bars or strong spiral arms).

\subsection{The collapse time-scales}

The gas computed following the previous section collapses to fall onto
the equatorial plane forming the disc as a secondary structure.  The
gas infall from the halo is parametrised by $ f g_{H}$, where $g_{H}$
is the gas mass of the halo and $f$ is the infall rate, that is, the
inverse of the collapse time scale $\tau$.

The collapse time-scale for each galaxy depends on its total mass
through the expression: $\tau \propto M_{9}^{-1/2} T_{9}$
\citep*{gal84}, where $M_{9}$ is the total mass of the galaxy in
10$^{9}M_{\odot}$ (column 7  of Table~\ref{grid_m}) and $T_{9}$ is
its age, assumed to be 13.2 Gyr in all cases.  We calculate a
characteristic collapse time scale $\tau_{c}$ for each galaxy from the
ratio of its total mass, $M_{9,gal}$, and the MWG, $M_{9,MWG}$:

\begin{equation}
\tau_{c}=\tau_{c,MWG}(M_{9,gal}/M_{9,MWG})^{-1/2}
\label{tau}
\end{equation}

where $\tau_{c,MWG}$ is the collapse time scale at the characteristic
radial region for the MWG, $R_{c}=R_{opt}/2=6.5$ kpc.  This collapse
time scale ($\sim 4$ Gyr) was determined from the corresponding value
for the Solar Neighborhood, $\tau_{\odot} \sim 8$ Gyr.  This value for
$\tau_{\odot}$ is very similar to that found in other standard
galactic chemical evolution models \citep{pcb98,chia99,cha01} and it
is constrained by a large number of data, such as the [O/Fe] {\sl vs}
[Fe/H] relation for stars in the halo and the disc, the present infall
rate, \citep[0.7 M$_{\odot}$pc$^{-2}$Gyr$^{-1}$, see][]{mir89}, 
or the G-dwarf metallicity distribution \citep{fer92,pardi94,kot02},
all of them well reproduced in our model with this long scale to form
the disc  \citep[see][]{gav04}.  We would like to emphasize that the
characteristic collapse time-scale computed with equation ~\ref{tau}
is not a free-fall time ( $\leq 1$ Gyr for the MWG), being much longer
than that since it has been calculated through the calibration with
the Solar Neighbourhood collapse time scale $\tau_{\odot}$.

The mass which does not fall onto the disc will remain in the halo,
and yields a ratio $M_{halo}/M_{disc}$ for the baryonic component
which is also in agreement with observations.  The relative
normalization of the halo, thick and thin disc surface mass densities
\citep{san87} gives an approximated proportion of 1:22:200, which
implies that the halo surface mass density must be about 1/100 that of
the disc component, also in agreement with the ratio obtained by the
multiphase model.

\begin{figure}
\resizebox{\hsize}{!}{\includegraphics[angle=0]{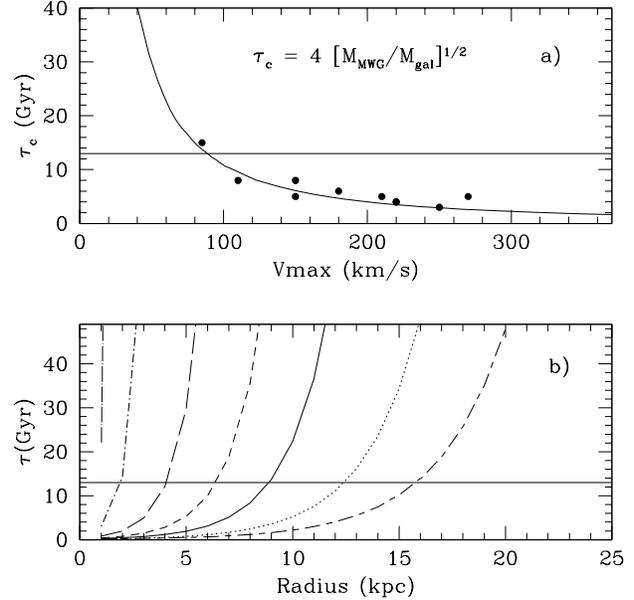}}
\caption{a) Characteristic collapse timescale $\tau_{c}$ for each
galaxy according to the maximum rotation velocity.  b) Radial
distribution of the collapse times scales $\tau_{coll}(R)$.  Lines
have the same meaning than in Fig.~\ref{dis} The horizontal solid line
represents the assumed age of galaxies, 13.2 Gyr.}
\label{tcoll}
\end{figure}

Fig.~\ref{tcoll} a) shows the characteristic time-scales {\sl vs} the
rotation velocity $\rm V_{opt}$ for our models. Solid dots represent
the values used in our previous models for individual spiral galaxies
\citep{fer94,mol96,mol99}.

An important consequence of the hypothesis linking the collapse time
scale with the total mass, is that low mass galaxies take more time to
form their discs, in apparent contradiction with the standard
hierarchical picture of galaxy formation.  This characteristic is,
however, essential to reproduce most observational features of spiral
and irregular galaxies \citep[see also][]{boi00}. In fact, recent
self-consistent hydrodynamical simulations in the context of a
cosmological model \citep{saiz04,dom04} show that a large proportion
of massive objects are formed at early times (high redshift) while the
formation of less massive ones is more extended in time, thus
simulating a modern version of the monolithic collapse scenario
\citep[][ELS]{ELS}. Furthermore, this assumption is supported by
various sets of observations
\citep{cim02,jim04,heav04,gla04,cim04,mc04} which seem to demonstrate
that a large proportion ($\sim 80$ \%) of the massive galaxies formed
their stars at $z> 1$ while a much smaller proportion of the less
massive ones have converted their baryon mass into stars at that
redshift.

It is evident that the collapse time-scale must vary with
galactocentric radius. Assuming that the total mass surface density
follows the surface brightness exponential shape, the required
collapse time-scale should also depend exponentially on radius, with a
scale length $ \lambda_{D} \propto R_{D}$ (where $\rm R_{D}$ 
is the corresponding
scale length of the surface brightness radial distribution). Thus, we
assume:

\begin{equation}
\tau_{coll}(\rm R)=\tau_{c}\exp{((\rm R-R_{c})/\lambda_{D})}
\label{tau_r}
\end{equation}

In principle, we might expect that $ \lambda_{D}=2
R_{D}$. However, in \citet{fer94} we have shown that such large scale
length for the infall rate produces a final radial variation for the
elemental abundances in disagreement with the observed distributions
in spiral galaxies.  \citep[see][ for a wide discussion about the
effect of this parameter on the radial distribution of
abundances]{por99}.  On the other hand, we must bear in mind that the
surface brightness distribution is the final result of the combination
of both the collapse and the star formation processes, and therefore
the collapse time-scale may have in principle a different dependence
on radius than the surface brightness itself.  For the sake of
simplicity, we assume a scale length $\lambda_{D} =0.15 R_{opt}$,
corresponding to half the scale length of the exponential disc
$R_{D}=R_{opt}/3.2$ given by PSS96.  The value of $\lambda_{D}$
decreases with the mass of the galaxy, in agreement with observations
\citep[][see their Fig.~4]{vau,guz96,grah01}.

 The radial dependence of the infall rate is not imposed {\sl a priori}
in our scenario, but it is consequence of the gravitational law and 
the total mass distribution in the protogalaxy. The physical meaning
is clear: galaxies begin to form their inner regions before the outer
ones in a classical inside-out scheme. This halo-disc connection is
crucial for the understanding of the evolution of a galaxy from early
times, the inside-out scenario being essential to reproduce the
radial gradient of abundances \citep{por99,boi00}. In fact, in a
chemo-dynamical model \citep*{sam97}, this scenario is produced
naturally.

The radial variation of the collapse time-scale for each galaxy,
calculated with equation \ref{tau_r}, is shown in Fig.~\ref{tcoll}b),
where we also draw a solid line at  13 Gyr, the assumed age of
galaxies. If the collapse time-scale is larger than this value, there
is not enough time for all the gas to fall onto the disc: only a small
part of it has moved from the halo to the equatorial plane and the
disc formation is not yet complete. This could correspond to the
situation observed by \citet*{san01}, who have found an extended
component of H\,{\sc i}, different from the cold disc, located in the
halo, rotating more slowly than the disc and with radial inward
motion.

\subsection{The star and cloud formation efficiencies}

The model computes the time evolution of each population which
inhabits the galaxy.  In the various regions of the disc or bulge and
in the halo, which are treated separately, we allow for different
phases of matter aggregation: diffuse gas ($ g$), clouds ($ c$, except
in the halo), low-mass stars ($s_{1}, m < 4 M_{\odot}$), high-mass
stars ($s_{2}, m \geq 4 M_{\odot}$), and stellar remnants
\footnote {All these quantities represent the total mass of each
phase, included in the code in units of $10^{9}M_{\odot}$}. The
value for stellar mass range division is related to nucleosynthesis
prescriptions: stars with masses lower than 4 M$_{\odot}$ only produce
light elements, and do not contribute to the interstellar medium
enrichment.\footnote{Also, this splitting of stars into two groups,
less and more massive than 4 M$_{\odot}$, allows a very easy
comparison of our resulting metallicity distribution with the observed
one, based on G-dwarf low mass stars.}

The mass in the different phases of each region changes by the
following conversion processes, related to the star formation and
death:

\begin{enumerate}
\item Star formation by the gas-spontaneous fragmentation in the halo
\item Cloud formation in the disc from diffuse gas
\item Star  formation in the disc from cloud-cloud collisions
\item Induced Star formation in the disc {\sl via} massive star-cloud
interactions
\item Diffuse gas restitution from these cloud and star formation
processes
\end{enumerate}

In the halo, devoid of molecular clouds, the star formation follows
a Schmidt law for the diffuse gas $g_{H}$ with a power $n=1.5$ and a
proportionality factor $K$. In the disk, stars form in two steps:
first, molecular clouds, $c_{D}$, form out of the diffuse gas,
$g_{D}$, also by a Schmidt law with $n=1.5$ and a proportionality
factor called $\mu$. Then, cloud-cloud collisions produce stars by a
spontaneous process at a rate proportional to a parameter
$H$. Moreover, a stimulated star formation process, proportional to a
parameter $a$, is assumed.

An advantage of using our model is that it includes a more realistic
star formation than classic chemical galactic evolution ones in
which the SFR prescriptions are based on a Schmidt law, depending on
the total gas surface density. Instead, the multiphase model assumes a
star formation which takes place in two-steps: first, the formation of
molecular clouds; then the formation of stars.  This simulates a power
law for the gas density with an exponent $n> 1$ and with a threshold
gas density as shown by \citet{ken89}, and, more importantly, it
allows the calculation of the two different gas phases present in the
interstellar medium. In fact, the actual process of star formation,
born by observations \citep{kle01}, is closer to our scenario with
stars forming in regions where there are molecular clouds, than to the
classical Schmidt law which depends only on the total gas density.

Our assumed SFR implies that some feedback mechanisms are included
naturally and are sufficient to simulate the actually observed process
of creation of stars from the interstellar medium. The formed massive
stars induce the creation of new ones. But, at the same time,
these star formation processes also may destroy the diffuse or
molecular clouds, thus preventing the total conversion of gas into
stars and ejecting more gas once again into the ISM --point (v)--.
In particular, massive stars destroy the molecular clouds that
surround them, due to the sensitivity of molecular cloud condensation
to the UV radiation \citep{par90}. This mechanism restores gas to the
ISM, thus decreasing the star formation.  Both regulating process are
included in our model, although neither heating or cooling mechanisms
for the cloud components are included explicitly in our code.

The complete set of equations is given in \citet{mol96} for each radial
region.  We summarize here only those related to the star formation
processes in the halo, $\Psi_{H}$, and in the disk, $\Psi_{D}$:
\begin{eqnarray}
\Psi_{H}          & = & Kg_{H}^{1.5}\\
\frac{dc_{D}}{dt} & = & \mu g_{D}^{1.5}-a c_{D}s_{D,2} -H c_{D}^{2}\\
\Psi_{D}          & = & H c_{D}^{2} + a c_{D} s_{D,2}
\end{eqnarray}
where H and D indicate halo and disk.  $K$, $\mu$, $H$ and $a$ ,
besides the previously described $\tau$ are the parameters of the model.

In the classical method of application of a chemical evolution model
to a given region or galaxy, the input parameters are considered as
free and chosen as the best ones in order to reproduce the selected
observational constraints of the galaxy. In our models, however, not
all these input parameters can be considered as free. The parameter
$f$ is defined by the total mass radial distribution, as already
explained. Regarding $K$, $\mu$, $H$ and $a$, we have tried to reduce
to a minimum their degree of freedom. Their radial dependence may be
estimated as \citep[see]{fer94}:

\begin{eqnarray}
K   & = & \epsilon_{K}(G/V_{H})^{1/2} \\
\mu & = & \epsilon_{\mu} (G/V_{D})^{1/2} \\
H   & = & \epsilon_{H} (cte/V_{D}) \\
a   & = & \epsilon_{a} (G\rho_{c})^{1/2}/<m_{s_{2}}>
\end{eqnarray}
where G is the universal gravitational constant, $V_{H}$ and $V_{D}$
are the halo and disk volume of each radial region, \footnote{The
volume of the disc is calculated with a scale height of 0.2 kpc for
all galaxies; the volume of the halo for each concentric region is
computed through the expression: $\rm V_{halo}(R) = 2 R_{halo}
\sqrt{(1-\frac{R}{R_{halo}})^{2}}$.} $\rho_{c}$ is the average cloud
density and $<m_{s_{2}}>$ is the average mass of massive stars. The
proportionality factors are called efficiencies that represent
probabilities associated with these processes.

In this way, the free parameters $K$, $\mu$, $H$ and $a$,
proportionality constants in our star and cloud formation laws, but
variable for each radial region, are computed through the efficiencies,
which represent the efficiencies of star formation in the halo,
$\epsilon_{K}$, cloud formation, $\epsilon_{\mu}$, cloud---cloud
collision, $\epsilon_{H}$, and interaction of massive stars with
clouds, $\epsilon_{a}$, in the disc. Only a given efficiency for each
process must be selected for the whole galaxy, although the original
parameters, and its corresponding processes, maintain their radial
dependence.

The term associated to the induced star formation describes a local
process and, as a result, its coefficient $\epsilon_{a}$ is considered
independent of the galaxy modelled and the location in it. The term
$\epsilon_{K}$ is also assumed constant for all haloes. Therefore, for
all our 440 models, both efficiencies take the same values already
used in our previous model for the MWG.  Only the other two
efficiencies $\epsilon_{\mu}$ and $\epsilon_{H}$ are allowed to vary
for each galaxy, being characteristic for each one of
them. Within each galaxy $\epsilon_{H}$ and $\epsilon_{\mu}$
are independent of the position.

The fact that galaxies with the same gravitational potential or mass
but different morphological type or appearance exist, implies that the
evolution of a galaxy does not depend solely on gravitation, even if
this may be the most important factor, but also on certain dynamical
conditions. These conditions cannot be taken into account, obviously,
in a simple chemical evolution model, but may change the evolution of
a galaxy, (mostly the star formation rate through temperature
variations). We may consider them as included in our efficiencies to
form molecular clouds and stars, $\epsilon_{\mu}$ and $\epsilon_{H}$,
which are allowed to change from one galaxy to another.

We assume that $\epsilon_{\mu}$ and $\epsilon_{H}$ vary between 0 and
1, according with its efficiency meaning. In principle both
parameters should be allowed to vary independently from each other,
what would increase very much the number of models to be
calculated. Furthermore, some of those models will not be physically
possible. On the other hand, on the basis of our previous works, a
trend between both efficiencies seems to exist, increasing or
decreasing together for a given galaxy respect to the values
appropriate for the MWG model.  We have therefore studied the
observational data related to molecular cloud and star formation from
the available gas in order to check if a correlation between these two
processes exists.

To this aim we have used the data from \citet{you96} which correspond
to a large sample of galaxies and refer to atomic and molecular gas
masses and to luminosities in the IR band and in H$\alpha$ emission,
two well known indicators of the star formation rate.  With them, we
have tried to establish if some correlation appears among our
efficiencies, $\epsilon_{\mu}$ and $\epsilon_{H}$.  Since the modelled
star formation rate have two steps to form stars, we will analyse the
transformation of diffuse gas in molecular gas and of the conversion
of the molecular clouds into stars.

From equations (10) and (11), our parameters $\epsilon_{\mu}$ and
$\epsilon_{H}$, efficiencies of transforming the diffuse gas 
in molecular clouds and forming  stars out of them, can be written as:
\begin{eqnarray}
\epsilon_{\mu}&=&\mu \frac{{V}^{1/2}}{0.67}\\
\epsilon_{H}&=&H \frac{V}{0.3}
\end{eqnarray}
while from equations (7) and (8) we have:
\begin{eqnarray}
\mu & = & \frac{(dc/dt)+SFR}{g^{1.5}}\\
H & = & \frac{SFR}{c^{2}}
\end{eqnarray}  

In these expressions the subscript $D$ has been omitted, since in all
cases we refer to the disc \footnote{Actually, $\Psi=hc^{2}+ ac
s_{2} $ but the second term is much smaller than the first and
therefore we can approximate $H\sim\frac{\Psi}{c^{2}}$}.

We may approximate $\frac{dc}{dt}\sim\frac{c}{\Delta t}$, where $\Delta
t$ is the mean time necessary to transform the diffuse gas in a
molecular cloud, in units of $10^{7}yr$:
\begin{eqnarray}
\epsilon_{\mu} &= & \frac{(c/\Delta t)+SFR}{g^{1.5}}\frac{{V}^{1/2}}{0.67}\\ 
\epsilon_{H} &= & \frac{SFR}{c^{2}}\frac{V}{0.3}
\end{eqnarray}  
where $g$ and $c$ are the total mass in diffuse and molecular gas in
units of $10^{9}M_{\odot}$, and SFR is the star formation rate in
units of $10^{9}M_{\odot}/10^{7}yr$. Therefore:

\begin{eqnarray}
\log{\epsilon_{\mu}}&=&\log{((c/\Delta t)+SFR)}-1.5\log{g}+0.5\log{V}+0.18\\
\log{\epsilon_{H}}& =& \log{SFR}-2\log{c}+\log{V}+0.5
\end{eqnarray}

SFR is usually estimated from the $H\alpha$ luminosity  \citep{you96}:

\begin{equation}
SFR (10^{9}M_{\odot}/10^{7}yr)=L_{H_{\alpha}} 3.4 10^{-10}
\end{equation}

and the volume of the disc is
\begin{equation} 
V=0.2 \pi R^{2}
\end{equation}

where R is in kpc and $L_{H_{\alpha}}$ is in $L_{\odot}$.  We have all
data to estimate these efficiencies, except $\Delta t$. Recent estimates
\citep{prin01,ber04} for this cloud accumulation time scale give
values several times the gravitational contraction scale 
(which is $\sim$
1-4 Myr), that is $\Delta t= 10-20$ Myr, and probably smaller than 50
Myr.  The probable range is $[3.10^{6}--8. 10^{7}]$ yr.  We have given
three possible values: $1. 10^{6}$, $5. 10^{7}$ and $5. 10^{8}$ yr in
order to take into account other possible slower modes \citep{pal04}.

\begin{figure}
\resizebox{\hsize}{!}{\includegraphics[angle=0]{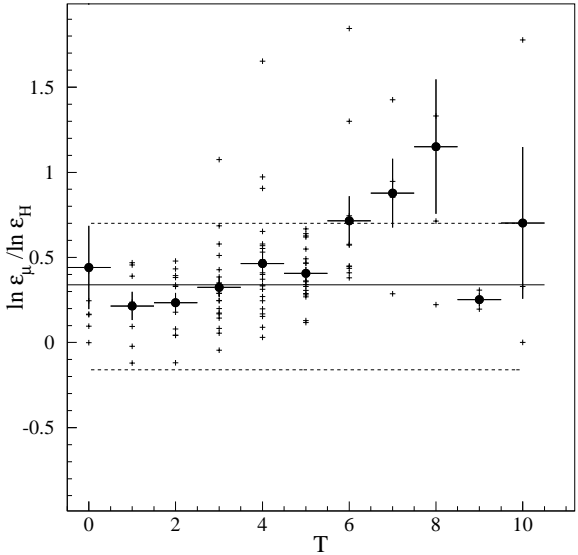}}
\caption{The ratio $ln \epsilon_{\mu}/ln \epsilon_{H}$ as a function
of the morphological type T. Solid dots are the averaged value for
each morphological type. The solid line is the mean value estimated
from these bins.}
\label{ratio}
\end{figure}

In Fig.~\ref{ratio} we represent the ratio $ln \epsilon_{\mu}/ln
\epsilon_{H}$ computed for $\Delta t= 5. 10^{7} yr$ as a function of
galaxy morphological type.  Single galaxies are represented as $+$
while solid dots correspond to average values obtained by binning them
into 11 types. A constant ratio for all types is consistent with the
data points, shown by the solid line and giving a mean value
$<\frac{ln \epsilon_{\mu}}{ln \epsilon_{H}}>\sim 0.34 (\pm 0.02)$ with
$\chi^{2}=3.6$. \footnote{The increasing ratio for increasing
type T is only apparent due to the reduced number of galaxies
within the bins in types from $T=8$ to 10.} 
Taking into account the large range
of variation of $\epsilon_{\mu}$ and $\epsilon_H$ (more than 5 orders
of magnitude) for the whole set of data, it is surprising that the $ln
\epsilon_{\mu}/ln \epsilon_{H}$ ratio takes values around 0.35 for all
galaxies.  For other values of $\Delta t$, this value changes but
remains constant. They are shown in the graph as dotted lines at
values 0.00 ($\Delta t= 1$ Myr) and 0.80 ($\Delta t= 50 $ Myr). 
Taking this into account,we have assumed a ratio $ln
\epsilon_{\mu}/ln \epsilon_{H}$= 0.4 for our computed models.

\begin{table}
\caption{Efficiencies chosen for our models. \label{efi}}
\begin{tabular}{ccc}
\hline
\noalign{\smallskip}
N & $\epsilon{\mu}$ & $\epsilon_{H}$ \\
\noalign{\smallskip}
\hline
\noalign{\smallskip}
1 & 0.95 &  0.88 \\
2 & 0.80 &  0.57 \\
3 & 0.65 &  0.34  \\
4 & 0.45 &  0.14  \\
5 & 0.30 &  0.05  \\
6 & 0.15 &  1.0e-2  \\
7 & 0.075 & 1.5e-3  \\
8 & 0.037 & 2.6e-4 \\
9 & 0.017 & 3.7e-5 \\
10& 0.007 & 4.0e-6 \\
\hline
\noalign{\smallskip}
\end{tabular}    
\end{table}

We have computed 10 models for each mass radial distribution, allowing
$\epsilon_{\mu}$ to take values between 0 to 1 as given in
Table~\ref{efi}.  For each one of these efficiencies,$\epsilon_{H}$
has been fixed according to the ratio $<ln \epsilon_{\mu}/ln
\epsilon_{H}> \sim 0.4$ and their values are also given in the table.
Each set of efficiencies has been labelled by a number N from 1 to 10.

{\sl Summarizing, only the characteristic collapse time-scale,
depending on the total mass, and the set of efficiencies, denoted by
number N, are varied from model to model.}  We thus obtain 440
different models which represent all possible combinations of the
collapse time-scale with the values of N.

\section{Results}

\subsection{The time evolution}

The results corresponding to the mass of each region and phase, the
star formation rate and the supernova rates, for the 440 computed
models are shown in Tables ~\ref{phases}, and ~\ref{sfrs}. The
elemental abundances for the discs are shown in Table
~\ref{abundances}.  Here we only give, as an example, the results of
the model corresponding to radial distribution number 22 and $N=5$,
with a rotation velocity of $\sim$ 140 km.s$^{-1}$, for the
first and last time steps of the evolution.

The whole tables with the complete time evolution from 0 to 13 Gyr,
with a time step of 0.5 Gyr, for the whole set of models will be
available in electronic form at CDS via anonymous ftp to
cdsarc.u-strasbg.fr (130.79.128.5) or via
{http://cdsweb.u-strasbg.fr/Abstract.html}, or
{http:/wwwae.ciemat.es/$^\sim$mercedes/grid}.

\footnotesize
\begin{table*}
\begin{minipage}{136mm}
\caption{Model Results corresponding to masses in each region 
and phase.}
\label{phases} 
\begin{tabular}{rrccccccc}
\hline
\noalign{\smallskip}
 Time & R & Mtot & Mdisc & Mgas(HI) & Mgas(H$_{2}$) & 
Mstars(M$\rm < 4M_{\odot}$) & Mstars(M$\rm \geq 4M_{\odot}$) & Mremnants \\
  (Gyr)  & (kpc) & (10$^{9}M_{\odot}$) & (10$^{9}M_{\odot}$) &
(10$^{9}M_{\odot}$) & (10$^{9}M_{\odot}$)  & 
(10$^{9}M_{\odot}$) & (10$^{9}M_{\odot}$) & (10$^{9}M_{\odot}$) \\
\hline
\noalign{\smallskip}
  ...  & ...& ...      & ...      & ...      & ...      & ...      & ...      & ...      \\
   0.1 &14. & 0.69E+01 & 0.47E--04 & 0.47E--04 & 0.89E--08 & 0.40E--17 & 0.31E--18 & 0.21E--19 \\
   0.1 &12. & 0.69E+01 & 0.22E--03 & 0.22E--03 & 0.11E--06 & 0.65E--15 & 0.49E--16 & 0.34E--17 \\
   0.1 &10. & 0.68E+01 & 0.10E--02 & 0.10E--02 & 0.14E--05 & 0.13E--12 & 0.93E--14 & 0.65E--15 \\
   0.1 & 8. & 0.67E+01 & 0.48E--02 & 0.47E--02 & 0.20E--04 & 0.28E--10 & 0.21E--11 & 0.14E--12 \\
   0.1 & 6. & 0.66E+01 & 0.22E--01 & 0.22E--01 & 0.31E--03 & 0.75E--08 & 0.56E--09 & 0.39E--10 \\
   0.1 & 4. & 0.58E+01 & 0.91E--01 & 0.87E--01 & 0.47E--02 & 0.21E--05 & 0.16E--06 & 0.11E--07 \\
   0.1 & 2. & 0.29E+01 & 0.21E+00 & 0.17E+00 & 0.39E--01 & 0.23E--03 & 0.17E--04 & 0.13E--05 \\
  ...  & ...& ...      & ...      & ...      & ...      & ...      & ...      & ...      \\
  13.2 &14. & 0.69E+01 & 0.58E--02 & 0.46E--02 & 0.13E--02 & 0.10E--04 & 0.22E--07 & 0.70E--06 \\
  13.2 &12. & 0.69E+01 & 0.27E--01 & 0.16E--01 & 0.10E--01 & 0.89E--03 & 0.16E--05 & 0.67E--04 \\
  13.2 &10. & 0.68E+01 & 0.13E+00 & 0.54E--01 & 0.43E--01 & 0.28E--01 & 0.31E--04 & 0.25E--02 \\
  13.2 & 8. & 0.67E+01 & 0.57E+00 & 0.14E+00 & 0.11E+00 & 0.29E+00 & 0.21E--03 & 0.32E--01 \\
  13.2 & 6. & 0.66E+01 & 0.23E+01 & 0.26E+00 & 0.19E+00 & 0.16E+01 & 0.79E--03 & 0.21E+00 \\
  13.2 & 4. & 0.58E+01 & 0.49E+01 & 0.18E+00 & 0.18E+00 & 0.39E+01 & 0.88E--03 & 0.65E+00 \\
  13.2 & 2. & 0.29E+01 & 0.29E+01 & 0.19E--01 & 0.51E--01 & 0.23E+01 & 0.95E--04 & 0.50E+00 \\
\hline
\noalign{\smallskip}
\end{tabular}
\end{minipage}
\end{table*}

\normalsize

In Table~\ref{phases} we list in column (1) the time, in Gyr and in
column (2), the galactocentric distance, in kpc, in Column (2).
Columns (3) to (9) give the mass, in units of 10$^{9} M_{\odot}$, in
each region and phase: column (3) the total mass in each region;
column (4) the mass of the disc region; column (5) the mass in the
diffuse gas phase;column (6) the molecular gas; column (7) the mass in
low and intermediate mass stars; column (8) the mass in massive stars
and, finally, column (9) the mass in remnants.

Table~\ref{sfrs} gives, for each time step in Gyr and radial distance
in kpc , listed in columns (1), and (2) respectively , the star
formation rate, in units of $M_{\odot} yr^{-1}$, in the disc and the
halo regions in columns (3) and (4); the supernova rates, for Types
Ia, Ib and and II, for the disc in columns (5), (6) and (7) and for
the halo in columns (8),(9) and (10) all of them in units of 100
yr$^{-1}$.

The abundances in the disc for 14 elements are given in
Table~\ref{abundances} for each time step in Gyr and radial distance
in kpc (columns (1), and (2) ).The abundances by mass of: H, D,
$^{3}$He, $^{4}$He, $^{12}$C, $^{13}$C, N, O, Ne, Mg, Si, S, Ca, and Fe
are given in columns (3) to (16).

The time evolution of several models is shown in the next figures.
In each one, 4 panels are shown, corresponding to 4 different maximum
rotation velocities and/or radial distribution of masses.  We have
selected the values corresponding to $\lambda=$0.03, 0.19, 1.0 and 2.5
corresponding to galaxies with rotation velocities of 48, 100, 200 and
290~ $\rm km s^{-1}$, respectively, and representing typical examples
of spiral and/or irregular galaxies.  For each panel we show the
results for the 10 selected values of the efficiencies (or
equivalently 10 rates of evolution) from $\rm N=1$, corresponding to
the highest efficiency values and hence the most evolved models, to
$\rm N=10$, with the smallest values and thus the least evolved ones.

\begin{figure*}
\resizebox{0.8\hsize}{!}{\includegraphics[angle=-90]{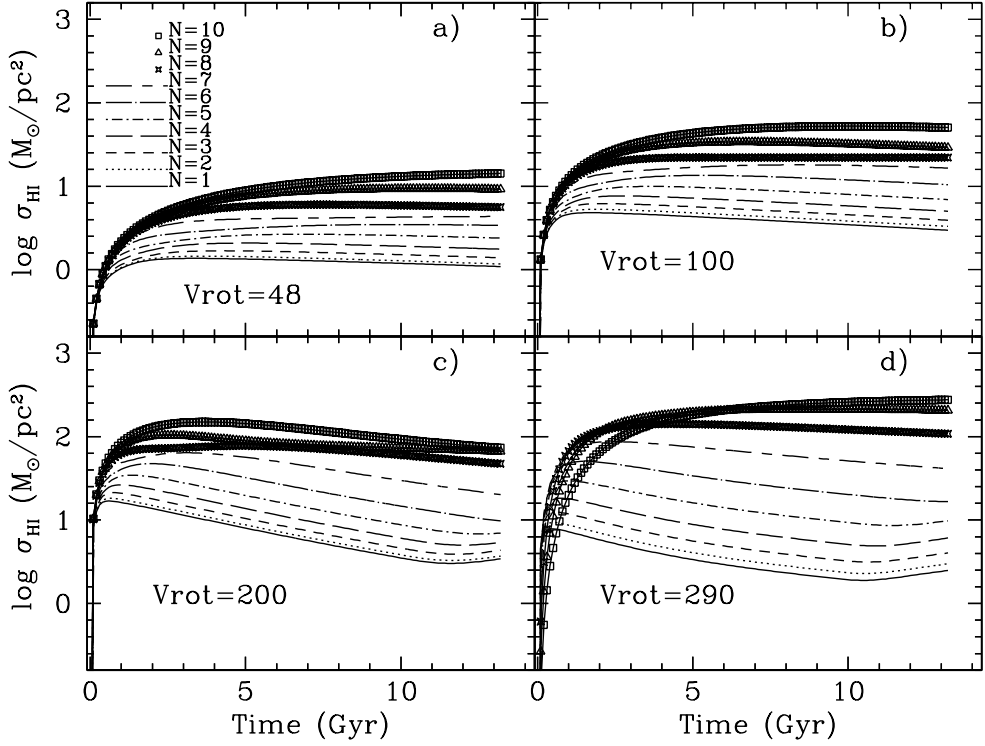}}
\caption{Time evolution of the logarithmic surface
density of the atomic gas of the radial region defined by $R_{c}$
for 4 different mass distributions following
the Vrot's values from each panel. In each one of them 10 models are
represented, for the 10 different efficiencies, following labels 
in panel a).}
\label{hit_rc}
\end{figure*}

We have computed models for galactocentric radii up
to $2.5R_{opt}$ with a step $\Delta R$ which depends on total galactic,
being larger (up to 4 kpc) for the most massive modelled galaxies and
smaller (only 1Kpc) for the lowest mass ones. Results are shown only
for the region
located at the galactocentric distance closest to the characteristic
radius $R_{c}$ defined in Table~\ref{grid_m}.

Fig.~\ref{hit_rc} shows the time evolution of the diffuse gas
density. In all cases, the diffuse gas surface density is seen to
increase rapidly at early times and then declines slowly. The first
abrupt increase is a consequence of the infall rate of gas from the
halo and hence does not depend on the model efficiencies. The increase
is faster for more massive galaxies. The later decline is due to gas
consumption in the process of molecular gas formation and hence
depends on the efficiency $\epsilon_{\mu}$, producing lower densities
for the galaxies with higher efficiencies.

\begin{figure*}
\resizebox{0.8\hsize}{!}{\includegraphics[angle=-90]{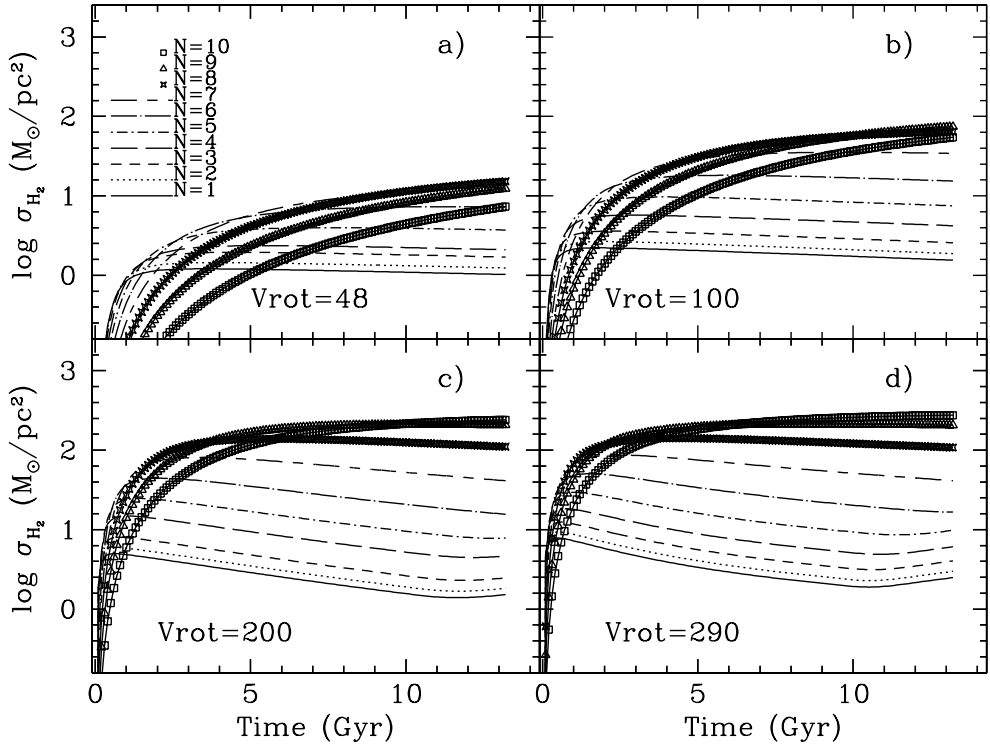}}
\caption{Time evolution of the logarithmic surface
density of the molecular gas of the radial region defined by $R_{c}$
for 4 different mass distributions following
the Vrot's values from each panel. In each one of them 10 models are
represented, for the 10 different efficiencies, following labels 
in panel a).}
\label{h2t_rc}
\end{figure*}

Fig.\ref{h2t_rc} shows the time evolution of the molecular gas phase
whose formation is delayed respect to that of the diffuse gas. It is
clear that the maximum density is reached later than that corresponding to the
diffuse gas density. For instance, for $\rm N=5$,
$\delta t=0.1$ Gyr for $\rm Vrot=290$ and 200 km s$^{-1}$, while it
is 0.3 Gyr for $\rm Vrot=100$ km s$^{-1}$ and reaches $\delta t=1.2$
Gyr when $\rm Vrot=48$ km s$^{-1}$. 

\footnotesize
\begin{table*}
\begin{minipage}{156mm}
\caption{Model Results: Star Formation Histories and Supernova Rates.}
\label{sfrs}
\begin{tabular}{rrrrrrrrrr}
\hline
\noalign{\smallskip}
Time  & R  & SFR(disc) & SFR(halo) & SN-Ia (disc) &  SN-Ib (disc) &SN-II (disc)& 
SN-Ia (halo) & SN-Ib (halo) & SN-II(halo)  \\
 (Gyr)  & (kpc) & (M$_{\odot}yr^{-1}$) & (M$_{\odot}yr^{-1}$) &
 ($100 yr^{-1}$) & ($100 yr^{-1}$) & ($100 yr^{-1}$) & 
 ($100 yr^{-1}$) & ($100 yr^{-1}$) & ($100 yr^{-1}$) \\
\hline
\noalign{\smallskip} 
  ...  & ...& ...      & ...      & ...      & ...      & ...      & ...      & ...      & ...      \\
   0.1 &14. & 0.26E--15 & 0.51E--01 & 0.17E--17 & 0.14E--18 & 0.42E--14 & 0.45E--02 & 0.11E--01 & 0.15E+01 \\
   0.1 &12. & 0.43E--13 & 0.52E--01 & 0.27E--15 & 0.14E--16 & 0.69E--12 & 0.46E--02 & 0.11E--01 & 0.16E+01 \\
   0.1 &10. & 0.82E--11 & 0.54E--01 & 0.52E--13 & 0.25E--14 & 0.13E--09 & 0.47E--02 & 0.12E--01 & 0.16E+01 \\
   0.1 & 8. & 0.18E--08 & 0.57E--01 & 0.12E--10 & 0.55E--12 & 0.30E--07 & 0.51E--02 & 0.12E--01 & 0.17E+01 \\
   0.1 & 6. & 0.49E--06 & 0.63E--01 & 0.31E--08 & 0.15E--09 & 0.80E--05 & 0.55E--02 & 0.14E--01 & 0.19E+01 \\
   0.1 & 4. & 0.13E--03 & 0.61E--01 & 0.89E--06 & 0.44E--07 & 0.22E--02 & 0.55E--02 & 0.14E--01 & 0.18E+01 \\
   0.1 & 2. & 0.14E--01 & 0.28E--01 & 0.11E--03 & 0.60E--05 & 0.24E+00 & 0.26E--02 & 0.65E--02 & 0.84E+00 \\
  ...  & ...& ...      & ...      & ...      & ...      & ...      & ...      & ...      & ...      \\
  13.2 &14. & 0.52E--05 & 0.45E--01 & 0.13E--05 & 0.56E--05 & 0.16E--03 & 0.18E--01 & 0.54E--01 & 0.14E+01 \\
  13.2 &12. & 0.38E--03 & 0.46E--01 & 0.10E--03 & 0.42E--03 & 0.11E--01 & 0.18E--01 & 0.54E--01 & 0.14E+01 \\
  13.2 &10. & 0.74E--02 & 0.46E--01 & 0.23E--02 & 0.86E--02 & 0.22E+00 & 0.18E--01 & 0.55E--01 & 0.14E+01 \\
  13.2 & 8. & 0.50E--01 & 0.44E--01 & 0.18E--01 & 0.59E--01 & 0.15E+01 & 0.18E--01 & 0.53E--01 & 0.13E+01 \\
  13.2 & 6. & 0.19E+00 & 0.29E--01 & 0.73E--01 & 0.22E+00 & 0.56E+01 & 0.13E--01 & 0.35E--01 & 0.86E+00 \\
  13.2 & 4. & 0.21E+00 & 0.25E--02 & 0.11E+00 & 0.25E+00 & 0.62E+01 & 0.24E--02 & 0.31E--02 & 0.74E--01 \\
  13.2 & 2. & 0.22E--01 & 0.17E--07 & 0.29E--01 & 0.28E--01 & 0.68E+00 & 0.14E--03 & 0.27E--07 & 0.51E--06 \\
\hline
\noalign{\smallskip}
\end{tabular}
\end{minipage}
\end{table*}
\normalsize

\begin{figure*}
\resizebox{0.8\hsize}{!}{\includegraphics[angle=-90]{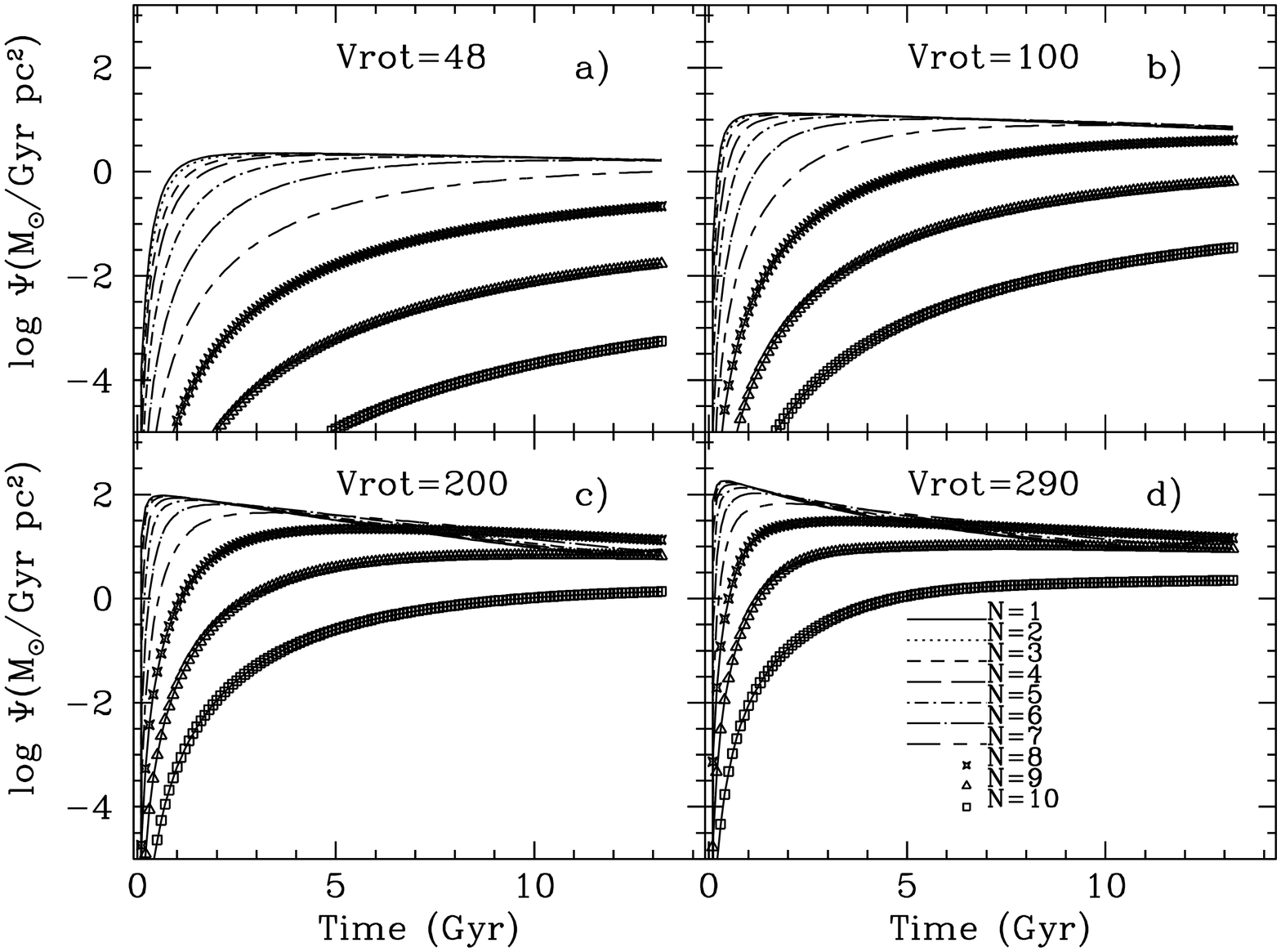}}
\caption{Time evolution of the logarithmic star formation rate surface
density of the radial region defined by $R_{c}$ for 4 different mass
distributions following the Vrot's values from each panel. In each one
of them 10 models are represented, for the 10 different efficiencies,
following labels in panel d).}
\label{sfrt_rc}
\end{figure*}

Fig.~\ref{sfrt_rc} shows the star formation rate history for the same
cases as before. These histories result extraordinary different, even
for equal efficiencies. Taking into account that the radial region
shown is equivalent in all galaxies, it indicates that the primary agent
driving the time evolution of the SFR is the galaxy mass through the
collapse time scale.  On the other hand, the resulting star formation
history in most models, is smooth, showing values larger than 10$^{2}
(\rm M_{\odot} Gyr^{-1} pc^{-2})$ only at the early evolutionary
phases of the most massive galaxies with very high efficiencies.

\begin{table*}
\begin{minipage}{126mm}
\vbox to 220mm{\vfil Landscape Table 5 to go here.
\caption{Model Results: Elemental Abundances.}
\vfil}
\label{abundances}
\end{minipage}
\end{table*}
\normalsize

\begin{figure*}
\resizebox{0.8\hsize}{!}{\includegraphics[angle=-90]{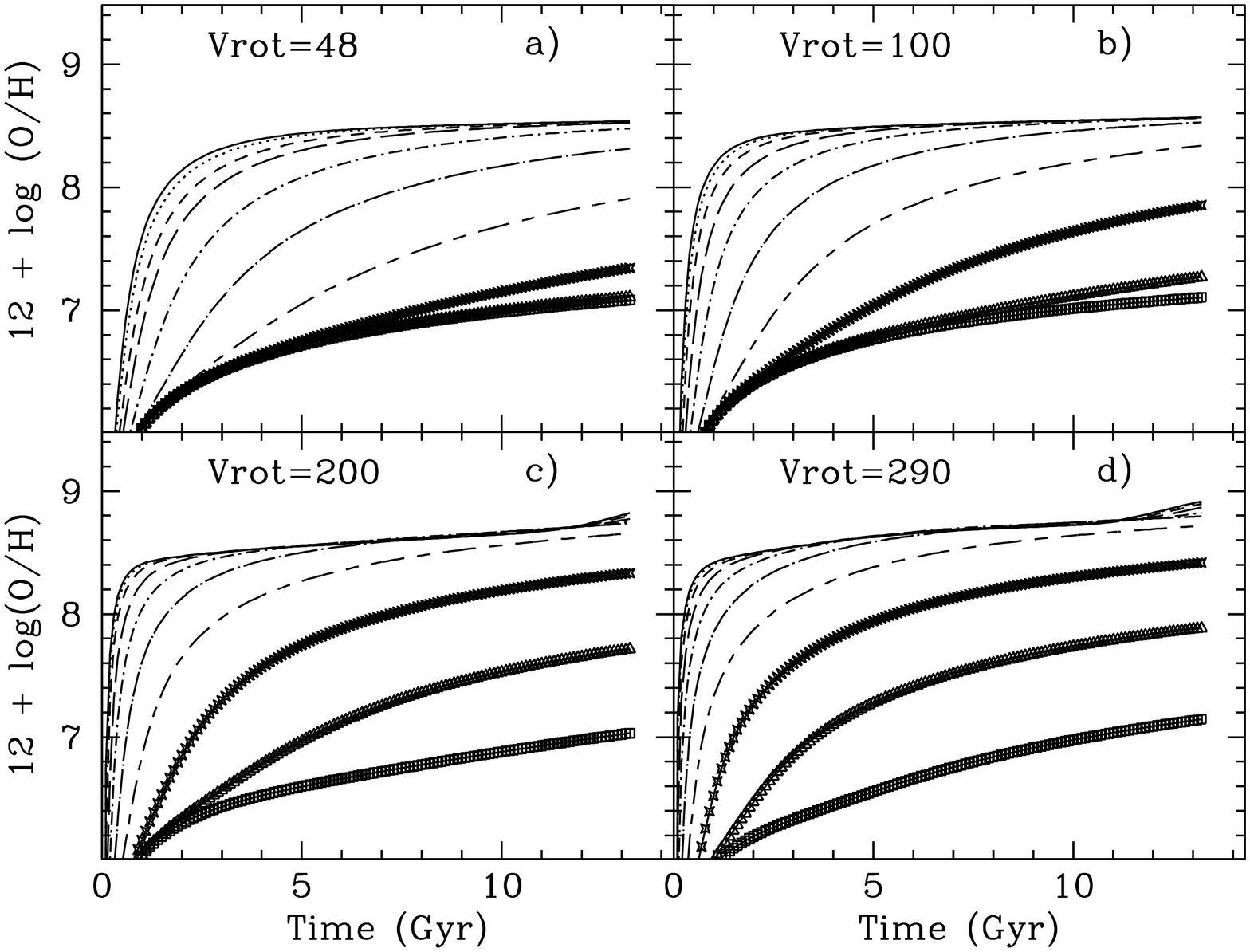}}
\caption{Time evolution of the oxygen abundance $12 + log(O/H)$ of the
radial region defined by $R_{c}$ for 4 different mass distributions
following the Vrot's values from each panel. In each one of them 10
models are represented, for the 10 different efficiencies, following
labels in panel d) of the previous figure.}
\label{oht_rc}
\end{figure*}

\begin{figure*}
\resizebox{0.8\hsize}{!}{\includegraphics[angle=-90]{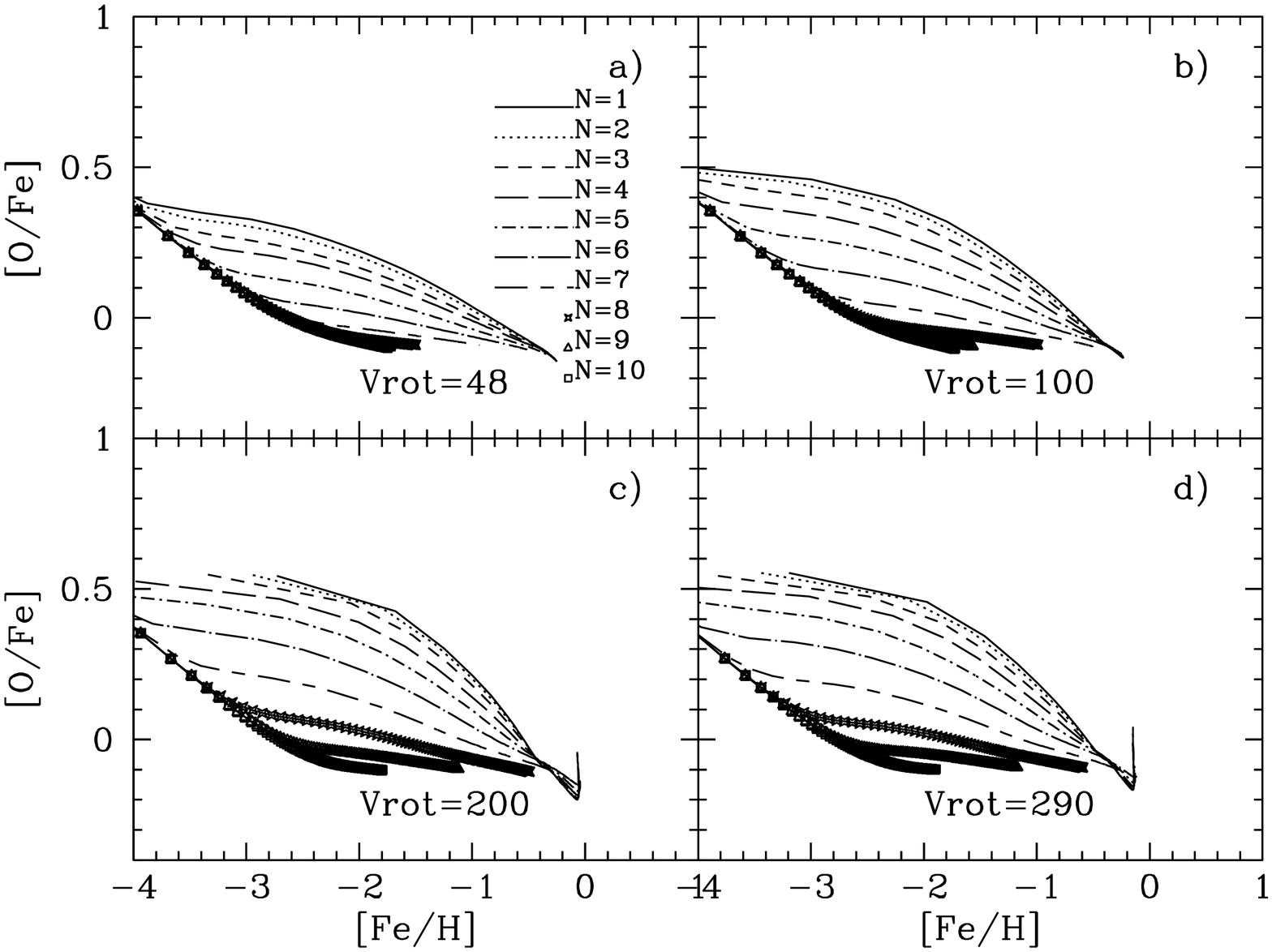}}
\caption{The evolution of the relative abundance [O/Fe]
{\sl vs} [Fe/H] for the radial region defined by $R_{c}$ for 4
different mass distributions following the Vrot's values from each
panel. In each one of them 10 models are represented, for the 10
different efficiencies, following labels in panel a)}
\label{ofe_rc}
\end{figure*}

The time evolution of the oxygen abundance relative to hydrogen,
expressed as 12+log(O/H), for the models described above is shown in
Fig.\ref{oht_rc}. Present time oxygen abundances look very similar for
all the galaxies with $N\leq5$. The models with low efficiencies,
$N\geq 7$, show very low abundances although always larger than
12+log(O/H)= 7, about the lowest observed oxygen abundance in
galaxies.

Fig.\ref{ofe_rc} shows the evolution of [O/Fe] when [Fe/H] increases.
The usual relation for the MWG can be taken to be represented by the
$N=4$ line from panel c), showing almost a {\sl plateau} for low
metallicities ($\rm [Fe/H]<1.5$) and then a decline toward the solar
value.  The {\sl plateau} appears in the models corresponding to
massive massive galaxies with high efficiencies \citep{mat98}. The
less massive galaxies show a slow decline without {\sl plateau} for
low values of N, and a steeper decrease for higher values. Thus, we
may expect that [O/Fe] in the less evolved galaxies will soon reach
the solar value \citep{mat92}.

\subsection{Present day radial distributions}

We now analyse the present time results obtained with this
bi-parametric grid of models. In order to do this, we start 
presenting only the present time radial distributions of gas, oxygen
abundances and star formation rate which we will analyse in the
following subsections. The radial distribution of gas and elemental
abundances provide two basic model constraints since they are easily
extracted from observations.  In addition, in many cases, the star
formation surface density and/or the surface brightness can also be
derived.

\subsubsection{Radial distributions of diffuse gas}

\begin{figure*}
\resizebox{\hsize}{!}{\includegraphics[angle=-90]{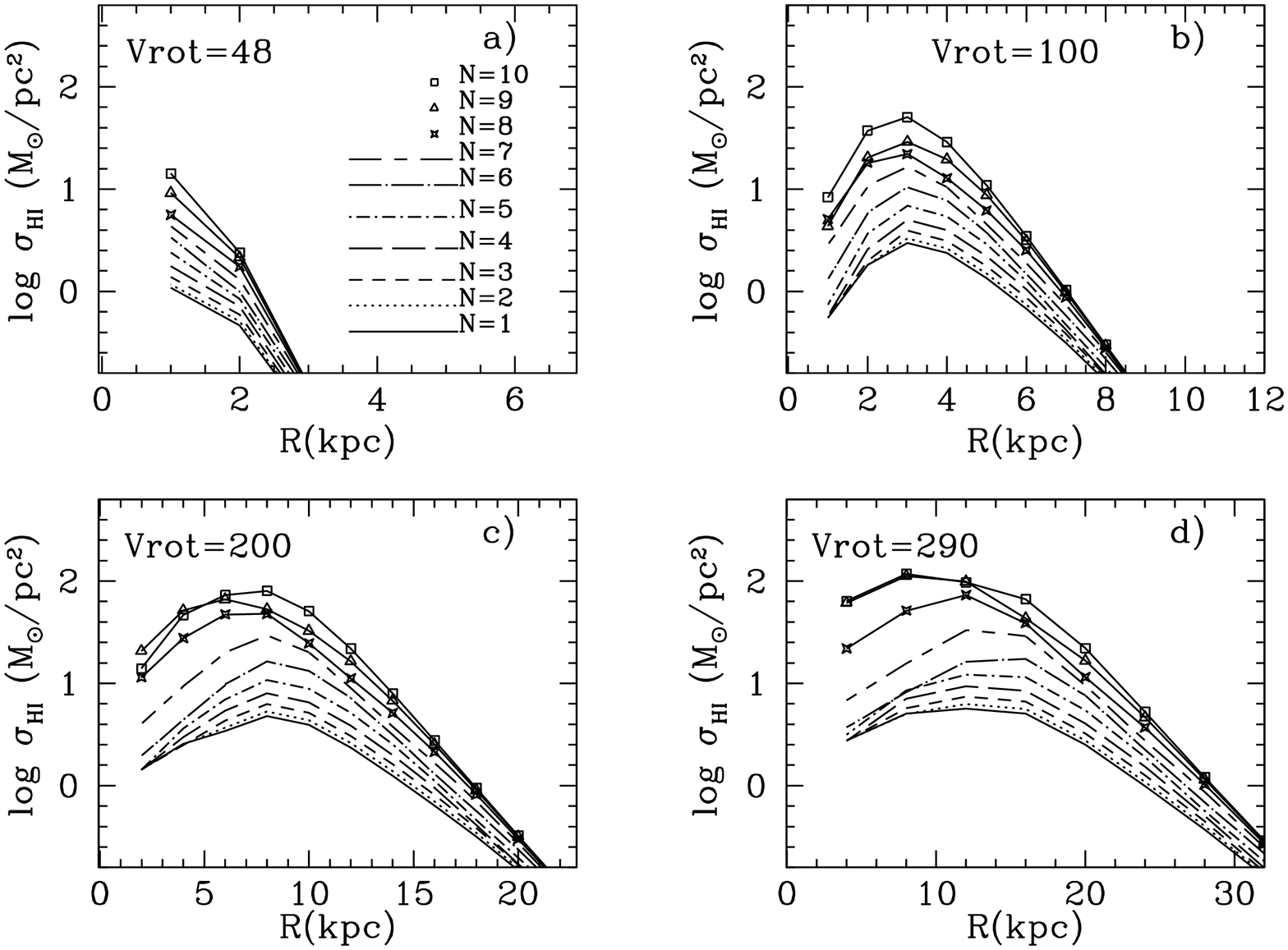}}
\caption{Present epoch radial distributions of the logarithmic surface
density of the atomic gas for 4 different mass distributions following
the Vrot's values from each panel. In each one of them 10 models are
represented, for the 10 different efficiencies, following labels 
in panel a).}
\label{hi}
\end{figure*}

The radial distribution of atomic gas surface density for the present
time is shown in Fig.~\ref{hi}. We can see that the atomic gas
surface density shows a maximum somewhere along the disc, as it is
usually observed \citep[][ and see
Fig.\ref{chequeo}]{cay90,rhee96,broe97}.  The value of this maximum
depends on N: the models with the highest efficiencies have smaller
gas quantities and their maximum densities are around 3-4
M$_{\odot}/pc^{2}$.  For intermediate efficiencies ($ 4 \leq N \leq
7$), these maximum values rise to $\sim 5-8$ M$_{\odot}/pc^{2}$. For
all efficiencies the radial distributions are very similar
independently of their galactic mass, except for those corresponding
to $\lambda\leq 0.03$ ($\rm Vrot= 48 km s^{-1}$) which show much lower
densities, except in the central region.

In fact, a characteristic shown by all distributions is the similarity
among models of the same efficiencies but different total mass, only
scaled by their different optical radius. With the exception of the
$\lambda=0.03$ model, all the others show, for a same N, differences
small enough as to simulate a dispersion of the data.

For the less evolved theoretical galaxies (N $ > 7$), the neutral gas
distribution display a clear dependence on galactic mass.  The maximum
density values are always large, as expected, due to the small
efficiencies to form molecular clouds, which do not allow the
consumption of the diffuse gas, but this maximum density increases
from $\rm \sim 15 M_{\odot}/pc^{2}$ for $\lambda=0.03$ ($\rm Vrot= 48
km s^{-1}$) to $\sim 30-40 M_{\odot}/pc^{2}$ for $\lambda=1.5$ ($\rm
Vrot= 248 km s^{-1}$) and up to $\sim 100 M_{\odot}/pc^{2}$ for
$\lambda=2.5$ ($\rm Vrot= 290 km s^{-1}$).

The consequence of a shorter collapse time-scale for the more massive
spirals is clearly seen: the maximum is located at radii further away
from the centre due to the exhaustion of the diffuse gas in the inner
disc which moves the star formation outside. The smaller the galaxy
mass, the closer to the centre is the maximum of the distribution,
which resembles an exponential, except for the inner region. In fact,
a shift in the maximum appears in each panel. In some cases, however,
the low values of the surface gas density are due to the fact that the
gas did not have enough time to fall completely onto the equatorial
disc. This effect is very clear for the model with $\lambda=0.03$
which shows a very steep distribution with gas densities lower than 15
$M_{\odot}/pc^{2}$.  In this case the gas shows a radial distribution
with a maximum at the centre. The low density out of this central region
is not due to the creation of stars but to the long collapse time-scale
which prevents the infall of the sufficient gas to be observed.
 
Besides the variations due to the differences in total mass, which
correspond to different collapse time-scales, and because we have
selected different efficiencies to form stars and molecular clouds, a
same total mass may have produced discs in different evolutionary
states.

Thus, a same $\lambda=0.15 $ may result in a disc of 7-8 kpc and
atomic gas densities around 5 $M_{\odot}/pc^{2}$, or a disc of only 5
kpc with a maximum density of 40 $M_{\odot}/pc^{2}$ in the region of 2
kpc.  In the same way, a galaxy with a large value of the total mass,
as the one with $\lambda=2.50$, may show a high gas mass density and a
little disc for a high value of N or, on the contrary, be very evolved
and therefore show no gas and a large stellar disc if the value of N
is low. The first object ($\rm N > 5$ ) could correspond to the
case of low surface brightness galaxies, while the last ones ($N < 4$) 
could be identified as the typical high surface brightness spiral
galaxies.

\subsubsection{Molecular gas radial distributions}

An important success of the multiphase models has been the ability to
reproduce the radial distributions for the atomic gas and the
molecular gas separately, which is possible due to the assumed star
formation prescription in two steps, thus allowing the formation of
molecular clouds prior to the appearance of stars. 
\begin{figure*}
\resizebox{\hsize}{!}{\includegraphics[angle=-90]{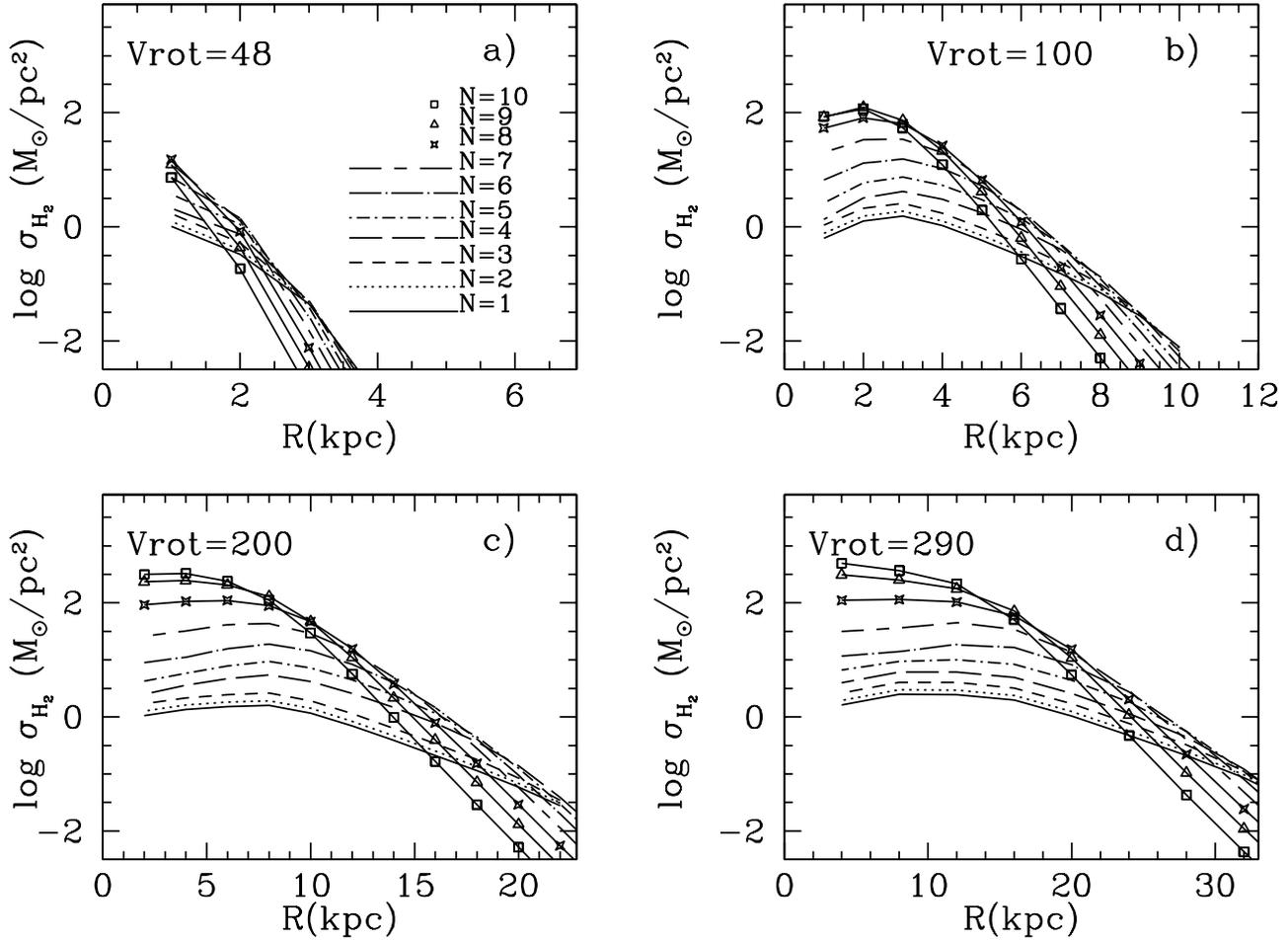}}
\caption{Same as Fig.\ref{hi} for the molecular gas surface densities.
Symbols meaning in panel a).}
\label{h2}
\end{figure*}

Another important consequence of this SFR law is that it takes into
account feedback mechanisms, even negative. If molecular clouds form
before stars, this implies a delay in the time of star formation.  The
molecular gas shows an evolution similar to that of the diffuse gas,
but with a certain delay. This delay allows the maintenance of a radial
distribution with an exponential shape, as usually observed, for a
longer time, although in some evolved galaxies $H_{2}$ is also
consumed in the most central regions, thus reproducing the so-called
central {\sl hole} of the molecular gas radial distribution, observed
in some galaxies \citep{nish01,reg01}, the MWG being one of them
(see Fig6b in next Section).

This is seen in Fig.\ref{h2} for efficiencies corresponding to $N>5$,
(depending on the total mass), for which the model lines turn over at
the inner disc, which corresponds to the regions located at the border
between bulge and disc.  Thus, the galaxies with the highest
efficiencies ($\rm N<5$) show a maximum in their radial distribution
of $\rm H_{2}$, which is always closer to the centre than that of the
atomic gas distribution.  Galaxies with the lowest efficiencies 
($N>5$) show larger surface densities of molecular than atomic gas
because the efficiency to form stars from molecular clouds is smaller
than the efficiency to form these clouds. Therefore, the conversion of
diffuse to molecular gas occurs more rapidly that the subsequent
formation of stars.

\subsubsection{The radial stellar discs profiles}

\begin{figure*}
\resizebox{\hsize}{!}{\includegraphics[angle=-90]{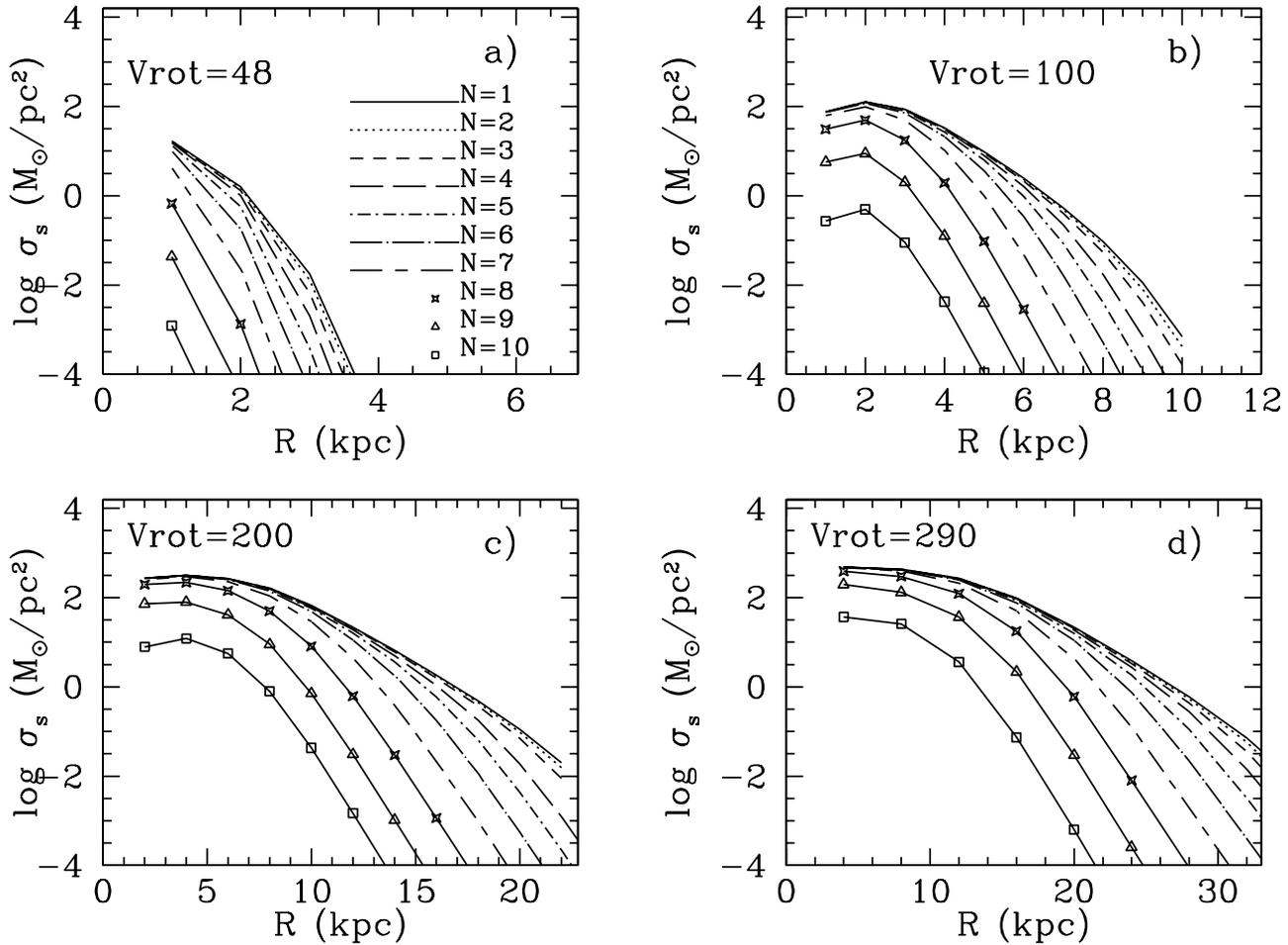}}
\caption{Present epoch radial distributions of total mass surface 
density for 4 different mass distributions. Symbols meaning in panel a).}
\label{profiles}
\end{figure*}

The total mass converted into stars forms out the stellar disc in each
galaxy.  These stellar discs are represented in Fig.\ref{profiles} as
the stellar surface density radial distributions.  These profiles may
be compared with brightness surface radial distributions, after the
appropriate conversions.

The values of efficiencies have a small influence in the resulting
stellar surface density distribution shape: the total mass of stars
created is similar for all N $< 6$, although they are formed at
different rates, that is the resulting stellar populations have
different mean ages. For the most evolved cases, most stars were
created very rapidly, while for the less evolved ones, stars formed
later as average, as it is shown in Fig.~\ref{sfrt_rc} for the
characteristic radius region.  Therefore, the radial distributions of
surface brightness would result very similar for galaxies of a given
galactic total mass, but colors are expected to be different, redder
for the galaxies with the highest efficiencies for the formation of
stars and molecular clouds.

A very interesting result is that the central value of the
distribution is practically the same, around 100 M$_{\odot}/pc^{2}$,
for all rotation curves and efficiencies, in agreement with Freeman's
law. Only galaxies with the smallest efficiencies or the less massive
discs show central densities smaller than this value. We cannot
compute a surface brightness only with these models, but assuming a
ratio $M/L=1 $ for the stellar populations, this implies a surface
luminosity density of $\rm \sim 100 L_{\odot}/pc^{2}$ and scale
lengths in agreement with observed generic trends. Only in models for
N$> 7$ the stellar discs show a different appearance: they look less
massive, as corresponding to discs in the process of formation.  This
implies that the surface brightness is lower than for the other types
for a similar characteristic total mass.

In any case, we remind that all information related to photometric
quantities, and this also applies to the disc scale lengths, which may
be obtained from the star formation histories and enrichment
relations of models, need the application of evolutionary synthesis
models, what is out of the scope of this work.

\subsubsection{The elemental abundances}

\begin{figure*}
\resizebox{\hsize}{!}{\includegraphics[angle=-90]{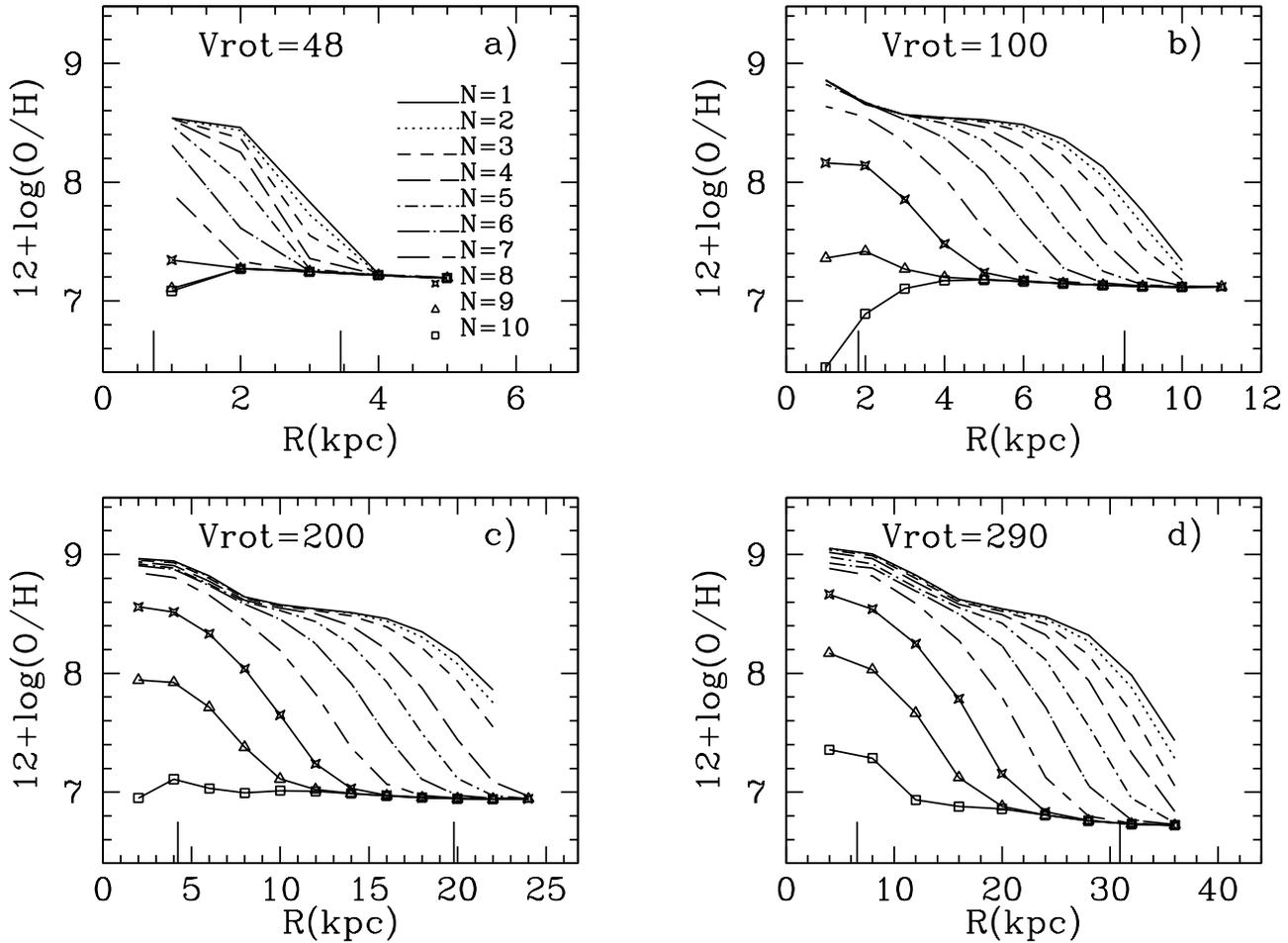}}
\caption{Present epoch radial distributions  of
Oxygen abundance, $12+log (O/H)$, for 4 different mass distributions.
Symbols meaning in panel a).}
\label{oh}
\end{figure*}

One of the most important results of this grid of models refers to the
oxygen abundances, shown in Fig.~\ref{oh}. A radial gradient appears
for most of models.  This is due to the different evolutionary rates
along radius: the inner regions evolve more rapidly that the outer
ones, thus steepening the radial gradient very soon for most model
galaxies. Then, the radial gradient flattens for the more massive
and/or most evolved (small N) galaxies due to the rapid evolution,
even in the outer regions, which produces a large quantity of
elements, with the oxygen abundance reaching a saturation level. This
level is found to be around $12+log(O/H) \sim 9.0- 9.1$
dex. Observations in the inner disc of our Galaxy support this
statement \citep{sma01}. This result was already found in our previous
works.

Moreover, the larger the mass of the galaxy, the faster the effect: a
galaxy with $\rm V_{rot}=100$ km.s$^{-1}$ has a flat radial gradient for
efficiencies corresponding to N$=$ 1 or 2, while a galaxy with
V$_{rot}=200$ km.s$^{-1}$, shows a flat distribution for $N< 4$.  The
less massive galaxies maintain a steeper radial distribution of oxygen
for almost all efficiencies, with very similar values of the
gradients.

Nevertheless, for any galaxy mass, if $N> 7-8$ the radial abundances
distribution are flat. Thus, the less evolved galaxies show no
gradient, such it is observed in LSB galaxies \citep{blok96}, the
intermediate ones show steep gradients, and the most evolved galaxies
have, once again, flat abundance radial distributions. The largest
values of radial gradients correspond to the intermediate evolutionary
type galaxies, with the limiting N varying according to the total mass
of the galaxy. The more massive galaxies only show a significant
radial gradient if N$\leq 8$ while the less massive ones have a flat
gradient only if $\rm N \geq 9$ with the rest having very pronounced
radial gradients even for $N=1$.

For the evolved galaxies, the characteristic efficiencies are high for
all the disc, thus producing a high and early star formation in all
radial regions. In this case, the oxygen abundance reaches very soon a
saturation level, flattening the radial gradient developed at early
times of the evolution.  The characteristic oxygen abundance, measured
at R$_{c}$, is higher for the more massive galaxies and lower for the
less massive ones. However, this correlation is not apparent when the
central abundance is used, due to the existence of the saturation
level in the oxygen abundance, which produces a flattening of the
radial gradient in the inner disc, even for galaxies with intermediate
efficiencies. Actually, this saturation level would correspond to the
integrated yield of oxygen for a single stellar population. As oxygen
is ejected by the most massive stars of a generation of stars, it
appears rapidly in the ISM. It is therefore impossible to surpass this
level of abundance.

In fact, the oxygen abundance radial distribution shows sometimes a
bad fit to observations in the central parts of the discs, which give
frequently abundances larger than 9.10 dex. This absolute value
of the oxygen abundance is not reached in any case by the models.  All
the computations performed within the multiphase approach reach a
maximum $12 + log (O/H) \sim 9.10$ dex which no model can exceed. We
might think that uncertainties in the calculations of the stellar
yields elements, still very dependent on the assumed evolution of
stars, are a possible reason for this saturation level appears in
theoretical models.  However, low and intermediate mass stars, for
which yields show large variations among the different groups
\citep[see][and references therein]{gav04}, depending on the assumed
hypothesis in the calculation of the stellar evolution in the latest
stellar evolutionary phases, do not produce oxygen; these
calculations only affect to N and C abundances.  On the other hand,
the most important uncertainty for the production of oxygen in massive
stars refers to the strength of stellar winds responsible for stellar
mass loss.  This mass loss implies less production of oxygen and a
larger ejection of carbon. In our models we have used Woosley \&
Weaver's yields which do no include stellar winds. Therefore, from
this point of view, the modelled oxygen abundances must be considered
upper limits.

It should be recalled, however, that all oxygen data yielding values
larger than 9.1 dex have been obtained from observations of H{\sc ii}
regions where the electronic temperature could not be measured, and
hence the oxygen abundances have been derived through empirical
calibrations which are very uncertain in the high abundance
regime. Actually, the shape of the radial distribution of oxygen
changes depending on the calibration used \citep[e.g.][]{ken96}. The
suspicion that these abundances are overestimated at least by 0.2 dex
is very reasonable \citep{pil00,pil01}.  Abundance estimates in H{\sc
ii} regions historically considered as metal-rich, result to be
almost solar once the electronic temperature has been finally
measured \citep{dia00,cas02,ken03,gar04}.

\subsubsection{The star formation rate}

Radial distributions of star formation rate surface density show an
exponential shape in the outer disc, but a less clear one in the inner
regions, where some models show a distribution flatter than that of the
molecular gas, and even decreasing toward low values of the star
formation rate in the centre.  These SFR radial distributions are very
similar to those estimated by \citet{mar01} from H$\alpha$ fluxes,
as we show in Fig.\ref{chequeo} for a few galaxies.

\begin{figure*}
\resizebox{\hsize}{!}{\includegraphics[angle=-90]{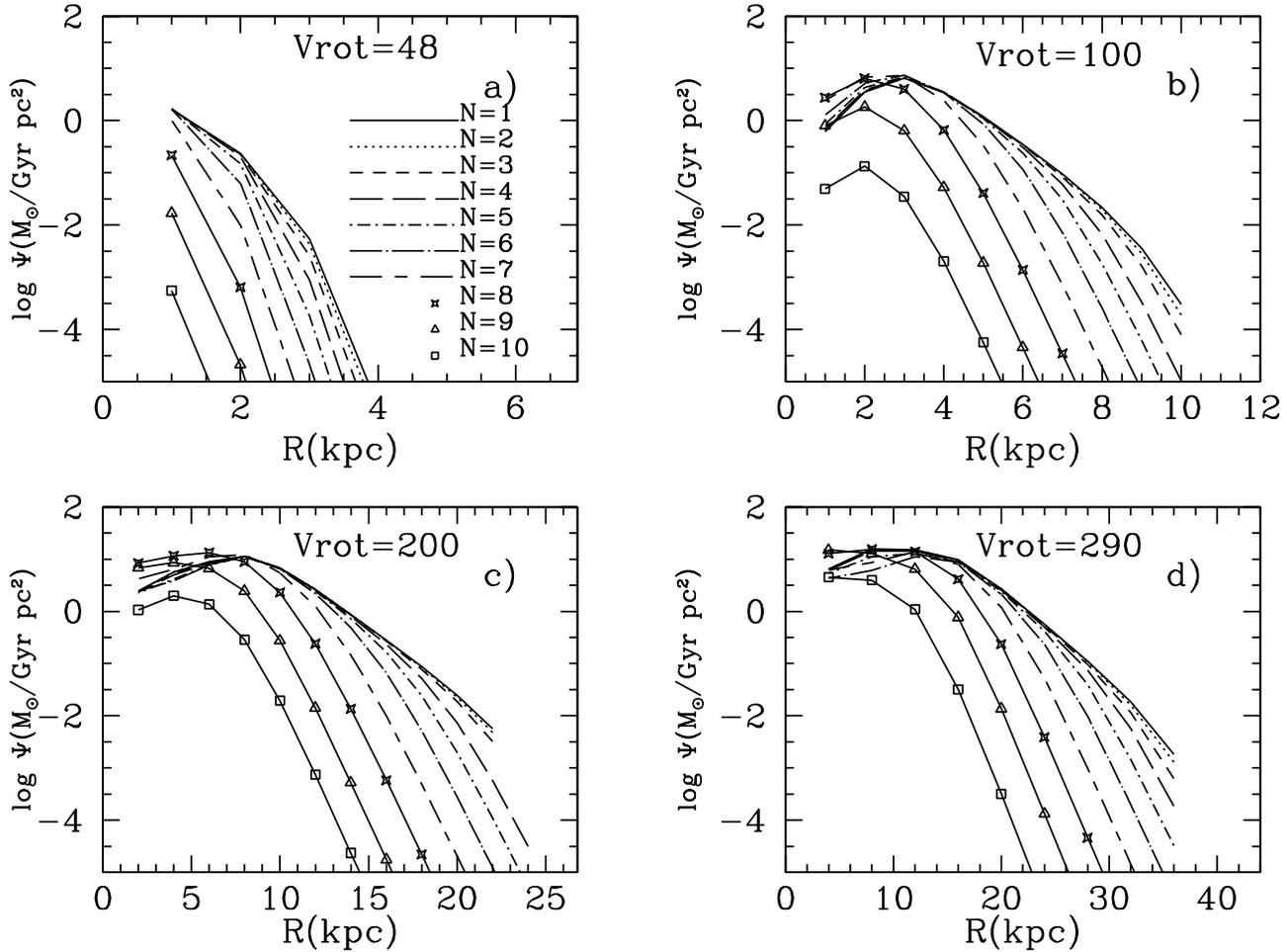}}
\caption{Present epoch radial distributions of the surface density of
the star formation rate  for 4 different mass distributions 
Symbols meaning in panel a).}
\label{sfr}
\end{figure*}

The star formation rate assumed in our models do not produce bursts of
massive stars in the low mass galaxies whatever the efficiencies.
Only the massive galaxies are able to keep a large quantity of gas in
a small region, usually at the centre although sometimes at the inner
disc regions. On the contrary, low mass galaxies collapse very slowly,
and thus the star formation rate maintains a low level during the
whole life of the galaxy. In fact, recent works suggest the same
scenario for both low mass and low surface brightness galaxies
\citep{dhoek00,leg00,bra01} in order to take into account the observed
data.  Our resulting abundances and gas fractions for low mass
galaxies seem to be in rough agreement with these findings
\citep{gav04b}, although our model results can not still be compared
with photometric data, for which many more observational sets exist.

\section{Discussion}

\subsection{Calibration with the MWG}

The first application of any theoretical model is to check its
validity for the MWG. A large set of observational data for the Solar
Neighbourhood and the galactic disc exists and therefore the number of
constraints is large compared to the number of free parameters of the
computed models. The model for $N=4$, and total mass distribution
number 28, corresponding to $\lambda=1.0$ and maximum rotation
velocity $V_{max}=200$ km s$^{-1}$, is the model more representative of
the MWG.

\begin{figure}
\resizebox{\hsize}{!}{\includegraphics[angle=0]{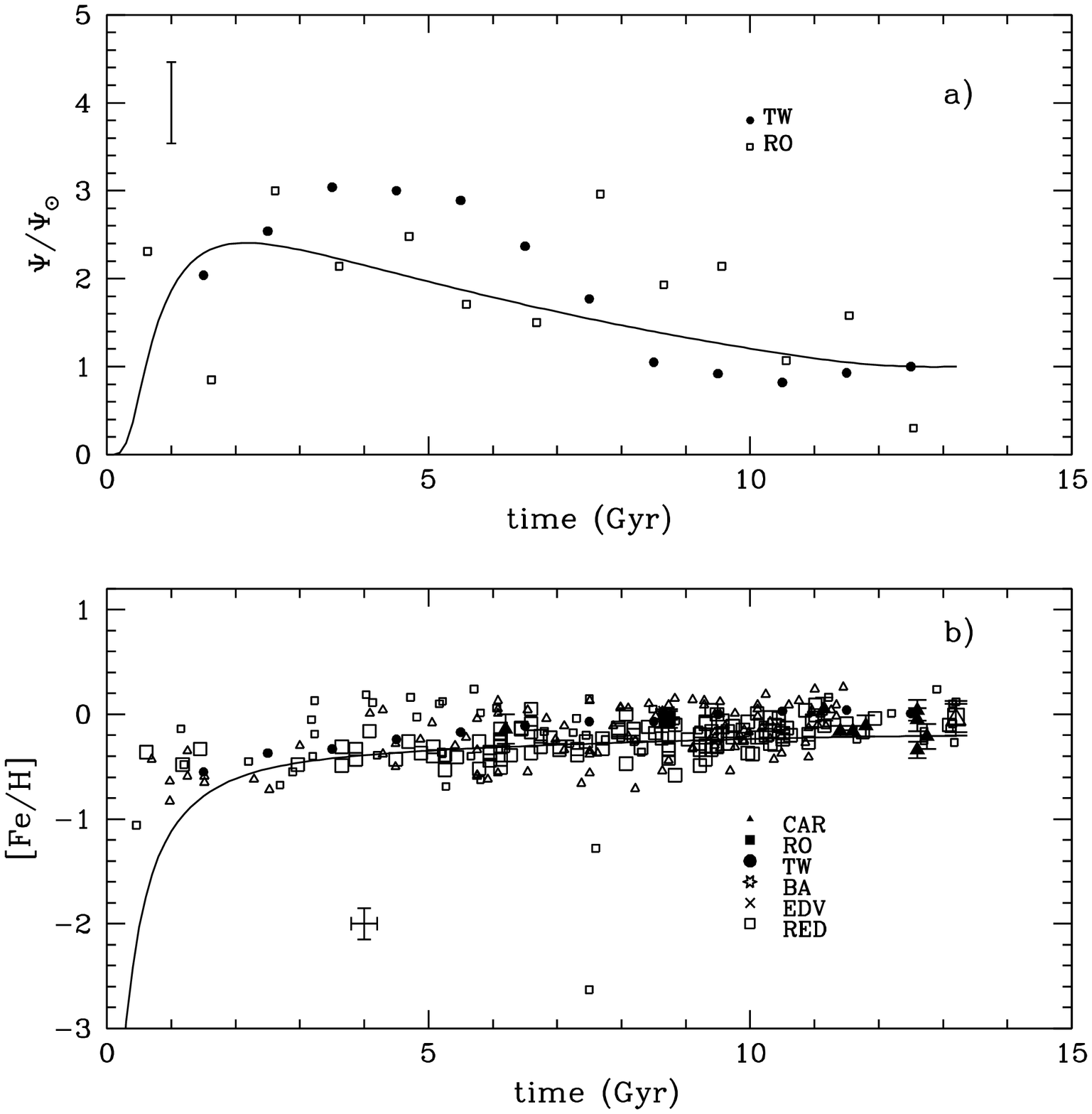}}
\caption{The Solar Neighborhood evolution as results from the chosen
MWG model for the region located at $R=8$ kpc. a) The star formation
history with data from \citet{twa80} --filled dots-- and
\citet{rocha00b} --open squares; b) The age-metallicity relation with
data from \citet{twa80}--TW--, \citet{ba88}--BA--, \citet{edv93}--EDV--, 
\citet{car98} --CAR--, \citet{rocha00} --RO-- and \citet{red03}--RED--, 
as labelled.}
\label{mwg1}
\end{figure}

\begin{figure*}
\resizebox{\hsize}{!}{\includegraphics[angle=0]{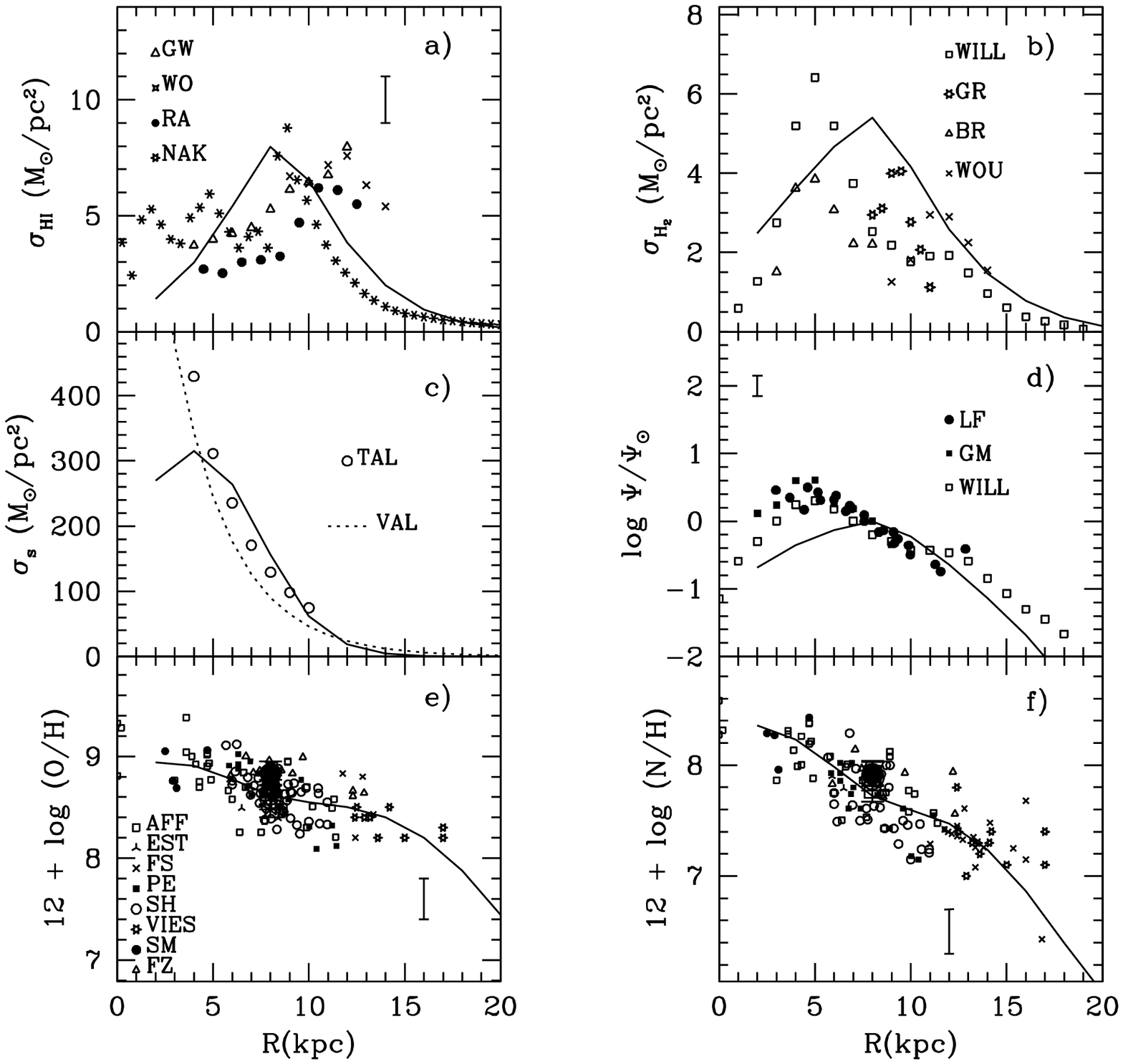}}
\caption{Present epoch radial distributions for the MWG simulated
galaxy (distribution number 28,$\lambda= 1.00$, $T=4$): a) atomic gas
density with data from \citet{gar89} --GW--, \citet{wou90} -WOU--,
\citet{rana91} --RA-- and \citet{nak04} --NAK; b) molecular gas
surface density with data from \citet{grab87} --GR, \citet{bron88}
--BR--, \citet{wou90} --WOU-- and \citet{will97} --WILL--; c) stellar
surface density with data from \citet{tal80}; the dashed line
corresponds to \citet{vall00}; d) star formation rate surface density
normalized to the present time solar value in logarithmic scale, data
taken from \citet{lac85,gus83,will97}, LF, GM and WILL, respectively;
e) and f) oxygen and nitrogen abundance as $\rm 12+log (X/H)$ with
data from \citet{aff97} --AFF-, \citet{este99,este99b,este99c}
--EST--, \citet{fich91} --FS--, \citet{fitz92} --FZ--, \citet{pei79}
--PE--, \citet{sha83} --SH-- , \citet{vil96}, --VIES-- and
\citet{sma97,sma01} --SM-- as labelled in panel e). Large filled
symbols at R$=8$ kpc in both panels represent the solar abundances
from \citet{gre98} -- circles--, \citet{hol01} --squares-- and
\citet{all01,all01b,all02}, --crosses--;  while the large empty symbols
are the interstellar medium abundances given by \citet{mey97,mey98},
--circles--, \citet{pei99} --squares--, \citet{sofia01}, --triangles,
and \citet{moos02}, --stars.}
\label{mwg2}
\end{figure*}

The results corresponding to this model are shown in Figs.~\ref{mwg1}
and ~\ref{mwg2} together with the available data.
Fig.~\ref{mwg1} shows the time evolution of the star formation history, 
panel a), and the age-metallicity relation, panel b), of the region
located at R$=$8 kpc from the Galactic centre, as compared to data for
the Solar Neighbourhood.  The observational trends are reproduced
adequately, although the model predicted maximum of the star
formation rate appears slightly displaced toward earlier times with
respect to observations.

Fig.~\ref{mwg2} shows the present time radial distributions of diffuse
and molecular gas, mass of stars, star formation rate and oxygen and
nitrogen abundances for the galactic disc.  The diffuse gas radial
distribution, panel a), is well reproduced in shape, although the
maximum observed gas surface density is somewhat displaced to outer
radii compared with the oldest data.  It, however, fits well the most
recent data obtained from \citet{nak04} for radii R$\ge 5$ kpc.
Regarding the molecular gas density, panel b), the distribution is
quasi-exponential from $R\simeq 8$ kpc and decreases at the inner disc
regions.  Taking into account that recent data give low densities at
these inner regions, we consider that our model results may be
adequate.  The stellar mass distribution, panel c), is exponential in
shape in agreement with the surface brightness distribution, and
radial distributions, panels e) and f), reproduce those shown by the
most recent data.

The only feature which is not well reproduced by the model is the SFR
radial distribution shown in panel d), which decreases toward the
inner disc in apparent discrepancy with observations. The modelled
star formation rate distribution has a maximum in $R\sim 7-8$ kpc,
while the observed one increases exponentially toward the galactic
centre, or levels off at 3-4 kpc.  On the other hand, the maximum of
the atomic gas density is observed around 10-11 kpc, and the molecular
gas density has its maximum observed at around 6 kpc. Therefore, it
results difficult to explain, from an observational point of view, how
the star formation rate remains so high at the inner disc (inside the
central 3-4 kpc), where both gas phases are already consumed. In fact,
the recent data from \citet{will97} show a decline for R$\le 5$ kpc
and  a decreasing star formation rate for the inner disc regions is
also observed in a large number of galaxies, as we will see in the next
section. Actually, the MWG radial distribution of the present star
formation rate is still a matter of discussion \citep{strong04} since
the detection of young sources, which are embedded in gas clouds, is
difficult and might be affected by selection effects. Data from pulsars,
supernovae or OB star formation \citep{bro00} seem to indicate that
the radial distribution of the SFR has a maximum around 5 kpc,
decreasing toward both smaller and larger radii. On the other hand
\citet{case96} obtained a much flatter distribution from the observed
cosmic rays and gamma radiation spectra. Since a decrease at the inner
disc is not unreasonable from the comparison with other spiral galaxy
data, we consider that the model results for the MWG are acceptable.

\subsection{Comparison with individual galaxies}

There is only a small sample of galaxies for which large observational
data sets, including neutral and molecular gas distributions, exist.
In what follows we show a comparison of our model results with the data
corresponding to theses galaxies.

\begin{table*}
\begin{minipage}{126mm}
\caption{Galaxy  Sample Characteristics and Model Input Parameters.}
\label{sample}
\begin{tabular}{lccccccccccc}
\hline
\noalign{\smallskip}
Galaxy & T & Type  & D & Vel$_{\rm rot,max}$  & Mass Distr.
& $\tau_{c}$ & R$_{c}$ &  N & $\tau_{old}$
& $\epsilon_{\mu,old}$ & $\epsilon_{H,old}$ \\
Name   &    & Class & (Mpc) & ($\rm km s^{-1}$)& Number 
& (Gyr) & (kpc) &  & (Gyr) & &      \\
\hline
\noalign{\smallskip}
NGC~300  & 7 & Scd/Sd  & 1.65 & 85  & 13 & 13.31 & 2.3 & 7& 13.3 & 0.07 & 0.007  \\
NGC~598  & 6 & Sc/Scd  & 0.84 & 110 & 21 &  8.13 & 2.9 & 6& 10.3 & 0.05 & 0.005 \\
NGC~628  & 5 & Sc      & 11.4 & 220 & 31 &  3.43 & 7.7 & 5&  3.28 & 0.25 & 0.01  \\
NGC~4535 & 5 & ---     & 16.6 & 210 & 30 &  3.59 & 7.4 & 5&  3.50 & 0.28 & 0.02  \\ 
NGC~6946 & 6 & Scd     & 7    & 180 & 25 &  4.94 & 6.2 & 6&  4.26 & 0.18 & 0.02  \\ 
\hline
\noalign{\smallskip}
\end{tabular}  
\end{minipage}
\end{table*}

The characteristics and corresponding input parameters of this galaxy
sample are given in Table~\ref{sample}. For each galaxy, Column (1),
the morphological type index is given in Column (2), while the
classical Hubble type is given in Column (3). The adopted distance
(taken from references following Table~\ref{data}, Column (2) is in
Column (4); the maximum rotation velocity is given in Column (5). The
number of the radial distribution of total mass --corresponding to the
column (1) of Table~\ref{grid_m}-- chosen to represent each galaxy is
in Columns (6).  The characteristic collapse time and the
characteristic radius corresponding to each distribution are given in
columns (7) and (8).  The number N chosen as the best one to reproduce
the observations is in column (9). The last columns, (10) to (12)
give, for comparison purposes, the collapse time-scale and
efficiencies for molecular cloud and star formation used in our
previous models \citep{mol96,mol99} for these same galaxies.

\begin{figure*}
\resizebox{\hsize}{!}{\includegraphics[angle=0]{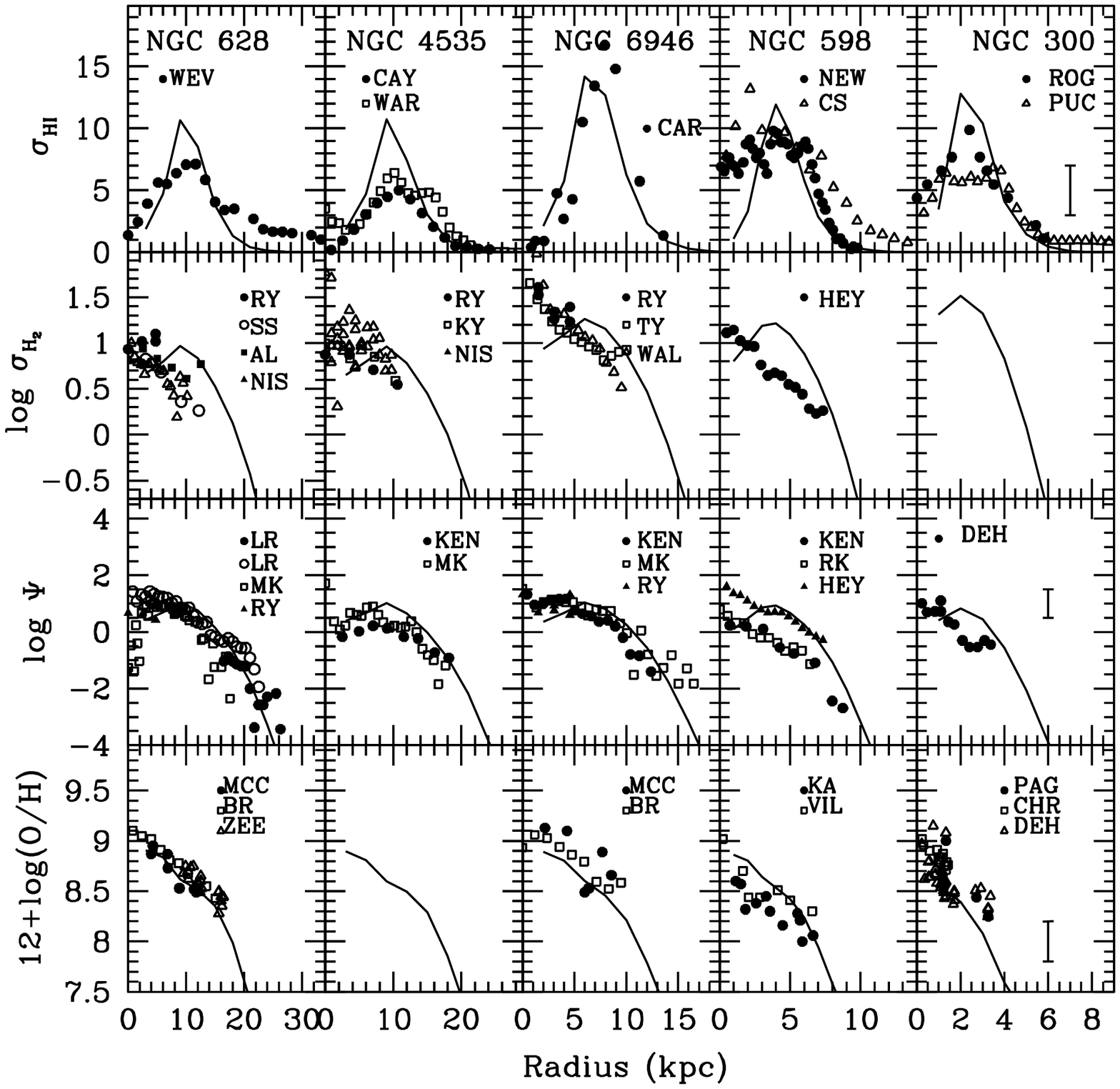}}
\caption{Present epoch radial distributions for atomic and molecular
gas densities, in units of $\rm M_{\odot}/yr$, (first and second
rows), star formation rate in $\rm M_{\odot} Gyr^{1} pc^{-2}$, (third row), and
oxygen abundance $\rm 12+log (O/H)$ (last row) for the sample used to
check the grid of chemical evolution models (Table~\ref{sample}).  The
observational data are taken from references given in
Table~\ref{data}.}
\label{chequeo}
\end{figure*}

The radial distributions of the different quantities: atomic and
molecular gas densities, star formation rate and oxygen abundance, for
these galaxies are shown in Fig.~\ref{chequeo} together with the
corresponding observational data, taken from the references given in
Table~\ref{data}.  

In the first row of panels of Fig.~\ref{chequeo} we can clearly see
that the radial distributions of neutral hydrogen are very well
reproduced by the models. For each galaxy, the distribution shows a
maximum along the disc, as predicted. Also, for galaxies with similar
total mass, such as NGC~4535 and NGC~6946, this maximum is higher for
later morphological type. For galaxies with the same value of T, but
different total mass, such as NGC~6946 and NGC~598, this maximum does
not change its absolute value, but it is located at a different
galactocentric distance, closer to the galaxy centre for the less
massive one, due to the longer collapse time-scale which reflects on a
slower evolution.  We would like to point out that the agreement
between model results and data is much improved when good quality data
(usually the most recently published ones) are used. The selection of
the best obtained distances improves extraordinarily this agreement,
as well. The same applies to the rest of the panels in this figure.

The radial distributions of molecular cloud surface density for each
galaxy, except NGC 300 for which no data exist, are shown in the
second row of Fig.~\ref{chequeo}. The agreement between model results
and observations is good for the outer discs which follow a quasi
exponential distribution, but is not so good for the inner discs where
the molecular hydrogen surface density is observed to continue
increasing instead of turn over as predicted by the models. We might
artificially decrease the efficiencies $\epsilon_{H}$ in these zones,
instead to maintain them constant as we do, and thus recover the
observed exponential shape. But in this case oxygen abundances will be
smaller than observed in these same regions. On the other hand we
should recall that the molecular masses are estimated from the CO
intensity through a calibration factor which depends on metallicity in
a way which would produce smaller molecular gas densities than usually
assumed at the inner galactic disc \citep{ver95,wil95}. Taking into
account that recent data yield low densities at these inner regions,
we consider that our model results may represent adequately the
reality. In any case, differences between models and data are larger
than in the case of the diffuse gas, which is not unexpected, given
the larger uncertainties involved in the derivation of molecular
hydrogen masses.

The radial distribution of the star formation rate for each galaxy is
shown in the third row of Fig~\ref{chequeo}. The agreement between
model results and data is very good for the most massive galaxies for
which both the maximum of the star formation rate and its location is
well reproduced. For the less massive galaxies, the predicted central
turnover of the distribution is not observed. 

In the last row of Fig.~\ref{chequeo}, the oxygen abundance
radial distribution for each galaxy, except NGC~4535 for which no data
exist, is shown.  The oxygen radial gradient is reproduced in all cases
and  also the observed trend of steeper radial
distributions -- larger radial gradients-- for the late type galaxies
is well reproduced by the models.

\begin{table*}
\begin{minipage}{116mm}
\caption{Galaxy Sample Data References.\label{data}}
\begin{tabular}{lccccc}
\hline
\noalign{\smallskip}
Galaxy  & D  & H\sc {i} & H$_{2}$  & SFR & [O/H] \\
\noalign{\smallskip}
\hline
\noalign{\smallskip}
NGC~300  & TU & ROG, PUC & ... & DEH  & PAG, CHR, DEH\\ 
NGC~598  & TP & NEW, CS & HEY & HEY, KEN, RK & KA, VIL \\
NGC~628  & MK & WEV & AL, NIS, RY, SS & LR, MK, RY & BR, MCC, ZEE \\
NGC~4535 & TP & CAY, WAR & KY, NIS, RY & KEN, MK  & ... \\
NGC~6946 & WE  & CAR & RY, TY, WAL & KEN, MK, RY & BR, MCC\\
\hline
\noalign{\smallskip}
\end{tabular}    
\footnotesize

AL: \citet{adl89}; BR: \citet{bel92}; CAR: \citet{car90}; 
CHR: \citet{chris97}; CAY: \cite{cay90}; CS: \citet{cor00}; 
DEG: \cite{degi84}; DEH: \citet{deh88}; HEY: \citet{hey04}; 
KY: \citet{kenn89}; KEN: \citet{ken89}; KA: \citet{kwi81}; 
LR: \citet{lr00}; MAR: \citet{mar01}; MCC: \citet{mcc85}; 
NN: \cite{nish01}; NEW: \citet{new80}; PAG: \citet{pag79}; 
PUC: \citet{puc90}; ROG: \citet{rog79}; RK: \citet{rum83}; 
RY: \citet{rownd99}; SS: \citet{sage89}; TY: \citet{ty89}; 
TU: \citet{tul88}; TP: \citet{tul00}; ZEE: \citet{vzee98}; 
VIL: \citet{vil88}; WAL: \citet{walsh02}; WAR: \citet{war88}; 
WEV: \citet{wev86}
\end{minipage}
\end{table*}

We would like to emphasize that the models shown in this section have
not been computed specifically for each of the sample galaxies, as was
done in our previous works. The models shown in Fig.\ref{chequeo} were
selected among the 10 available ones for their corresponding rotation
velocity following Table\ref{dis} of the grid. Thus the good agreement
found between model results and data demonstrate that our bi-parametric
models are able to adequately reproduce real galaxies.

\subsection{Star formation rate {\it vs} gas surface density}

\begin{figure}
\resizebox{\hsize}{!}{\includegraphics[angle=0]{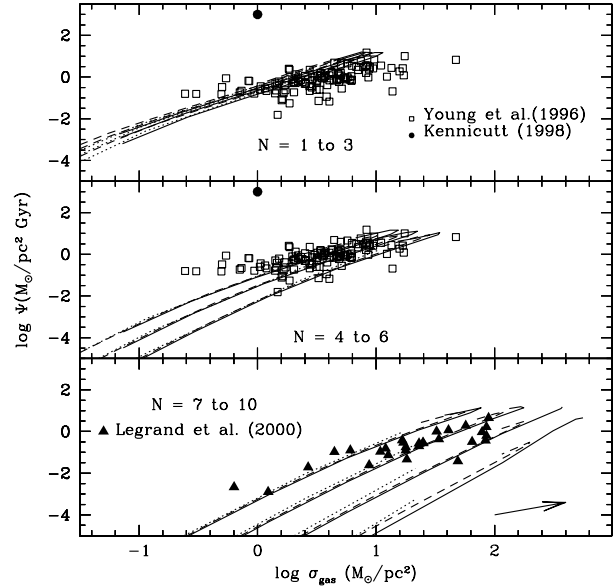}}
\caption{The relation of the surface density of the star formation
rate with the total gas density for 4 different mass distributions: a)
models with $N \le 3$; b) models with $4\le N \le 6$; and c) models
with $ N \ge 7$; Data are from \citet{you96} --open squares- and
\citet{ken98}--filled dots--, in panels a) and b), while those from
panel c) are given by \citet{leg00}.}
\label{ken}
\end{figure}

The computed star formation rate reproduces the relation obtained by
\citet{ken89}, when it is represented {\sl vs} the total gas surface
density as can be seen in Fig.~\ref{ken}.  In the two upper panels of
the figure ($ N \leq 6$), we have over-plotted the data from
\cite{you96} and \cite{ken98} which are seen to fall in the locus
defined by the models.  Models with $N\geq 7$ seem to be out of the
region where this data lie.  However, due to their extremely low
efficiencies, we may assume that these models do not simulate normal
bright spiral galaxies, but other kind of objects with a low stellar
content and high gas fractions more similar to dwarf and LSB galaxies.
We have, therefore, taken data from \cite{leg00} on these kinds of
objects and computed the densities for both quantities assuming an
optical radius of 5 kpc for all of them. Obviously, a change in this
radial dimension would vary the final values of our estimates, but our
hypothesis is probably valid within a factor of 2 (radii less than 10
kpc). Under this assumption, we see that the points, shown in the
third panel of Fig.\ref{ken}, fall in the upper locus defined by the
models.  If the radius were smaller than assumed, the densities would
be even higher, and the points would move in the figure following the
direction given by the arrow.  A second factor not included in these
estimates, is the molecular gas content. There are some works
suggesting that the molecular gas amounts to less than 10\% in this
kind of galaxies while other seems to indicate that the molecular mass
may be as large as in the brightest massive spirals.  In any case, the
inclusion of this factor would move the points to the right. In both
cases the data points would populate the region of the diagram
occupied by the models.  We therefore conclude that our models are
able to reproduce the observed trend of SFR {\sl vs} total gas surface
density for different types of galaxies.  

\subsection{Radial gradients of abundances}

The model computed abundance radial gradients for oxygen
have been obtained by fitting a least squares straigh
line to the oxygen abundances for radii
$\rm 0.5Rc kpc  \le R \le 3Rc$ kpc. The chosen radial range 
tries to eliminate the central region (where the oxygen abundance
distribution flattens) and the outer region where there are no
data. Thus the calculated gradients may be compared more
precisely with the corresponding observed ones.

\begin{figure}
\resizebox{\hsize}{!}{\includegraphics[angle=0]{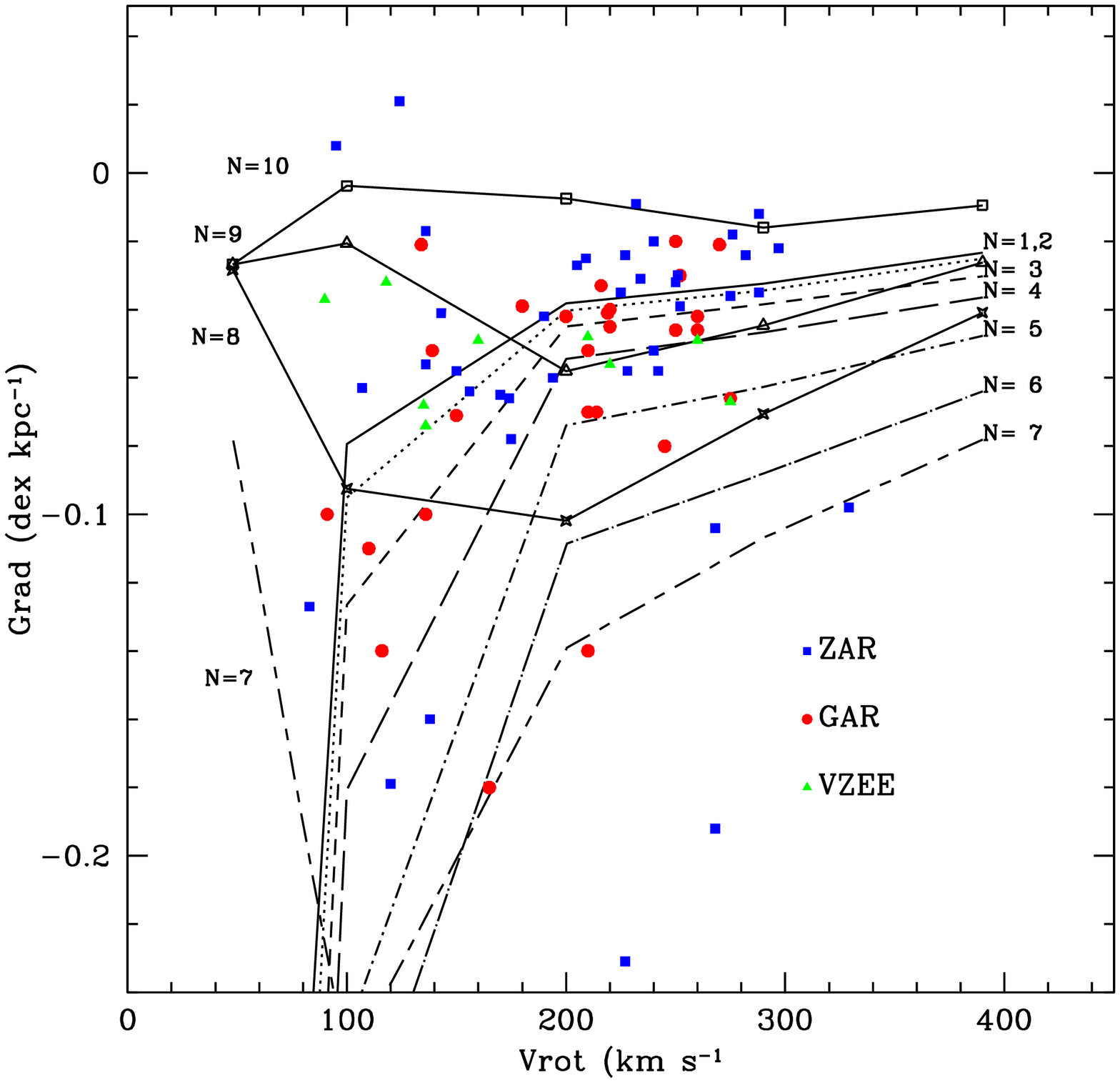}}
\caption{Radial gradients in dex kpc$^{-1}$ as obtained from the
modeled oxygen abundance radial distributions, computed with a least
square straigh line in the radial range $\rm 0.5Rc kpc < R < 3Rc kpc$,
as a function of the rotation velocity Vmax in kpc.s$^{-1}$. Models
are represented by the same line coding than Fig. 10 to 14. Data are
from \citet{zar94,gar97,vzee98}, shown by squares, dots and triangles,
respectively.}
\label{grad}
\end{figure}

These radial gradients, measured as dex/kpc, are represented as a
function of rotation velocity in Fig.~\ref{grad}.  The relation
between both quantities shows that the radial distributions are
steeper for the models with lower rotation velocities if N$\le 7$.
The radial gradients for a given rotation velocity are larger in
absolute value for increasing N, and tend to zero value for the most
massive galaxies and low N.  The models with $N > 7$, however, tend to
deviate from this function and show lower absolute values. That is,
for a similar rotation curve, their radial distributions of abundances
show a flattening as compared to those corresponding to the models
with $N \leq 7$ .  The rotation velocity at which the models deviate
from the common locus depends on $N$: for $N=7$ it occurs for
$V_{opt}\sim 100$ while for $N=10$ almost all rotation velocity curves
produce flat radial gradients of abundances.

Data from \citet{zar94,gar97,vzee98} are over-plotted on this
Fig.~\ref{grad}. We see that the modelled trend reproduce the
observations, although a more profound analysis about this (and other) 
correlation will be performed in the future.

\subsection{Efficiencies}

In order to use this grid for a given galaxy for which observational
data are known, we must first select the radial distribution of
mass. This means that we must know the total mass, the maximum
rotation velocity or, if none of them is available, the luminosity or
the magnitude in the I band.  According to this value we may choose
the number of the distribution from Table~\ref{grid_m}.  Then, 10
different models, corresponding to the 10 different efficiencies, are
available.

The standard procedure in chemical evolution would be to see which of
them is able to reproduce with success known data such as elemental
abundances or gas densities. If more than one observational constraint
is available, some kind of minimum error or maximum probability
technique may be used for the purpose of choosing the best model. Once
this selection is performed we may use the time evolution given by the
chosen model to predict the star formation history, the
age-metallicity relation, the stellar populations or any other
quantity relative to the modelled galaxy.  Obviously, the largest the
observational number of constraints, the smallest the uncertainty in
the selection of the best model.  The old and well known uniqueness
problem of the chemical evolution suggest that more than one model may
reproduce the data. Actually, we have shown in \citet{mol02} that this
problem reduces greatly when more than two observational constraints
can be used. Only 4\% of the 500 models computed with with different
input parameters could reproduce the observations relative to HI,
oxygen abundance and SFR at the same time. Therefore we are confident
that models may provide the evolutionary history of a given galaxy
within a reasonable accuracy.

Sometimes, however, these observational constraints are not available
(or not with sufficient precision) to select the adequate model.  In
that case, some other method to choose the possible evolutionary track
is necessary.  Our best models from previous works give us evidence
that the efficiencies to form molecular clouds from atomic gas and
stars from molecular gas, seem to depend on the galaxy morphological
type. If this were the case, the selection of the best model might be
easier. In fact, this dependence was already found in
\citet{fer88,gal89}, where these authors quantified the efficiencies
to form molecular clouds and the frequency of cloud-cloud collisions,
finding that a variation of 10 in the parameters H and $\mu$ is needed
when the Hubble type changes from one stage to the next (Sa to Sb,
etc..).

\begin{figure}
\resizebox{\hsize}{!}{\includegraphics[angle=0]{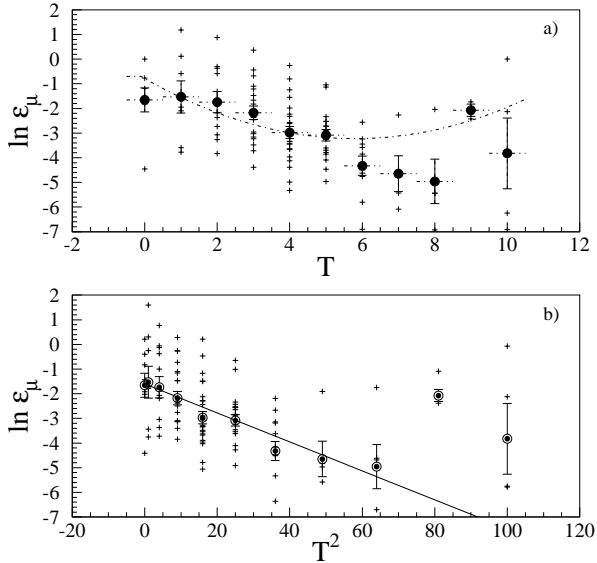}}
\caption{a) The efficiency $ln \epsilon_{\mu}$ as a function of the
morphological type T represented by $+$ for all galaxies. The solid
dots are the averaged values computed with 10 bins. The solid line is
a least-square second degree polynomial.  b) The efficiency $ln
\epsilon_{\mu}$ as a function of $T^{2}$. The solid line is the
least-squares straight line.}
\label{emu1}
\end{figure}

In order to check if a relation between efficiencies and
morphological type exists, we have plotted in Fig.~\ref{emu1}, panel
a) the logarithmic efficiency $\epsilon_{\mu}$, computed following the
equation (19) from section 2.3, as a function of T for the data of
\citet{you96}. There $+$ symbols correspond to the raw values while
solid dots represent the averaged values obtained for 10 bins, one for
each T.  A clear decreasing correlation appears for $T < 9$. Values
for $T\geq9$ fall above the trend but the number of points is small
for what we assume that the final increase is probably
spurious. Based on our previous works, we expect a relation of the
form: $\epsilon_{\mu} \propto exp^{-T{2}/A}$.  Therefore, in panel b)
we have plotted the logarithmic efficiency as a function of $T^2$
where a linear correlation is apparent.  A least-squares fit gives,
for $T < 9$ : 

\begin{equation}
<ln \epsilon_{\mu}>=-1.60-0.057 T^{2} , \chi^{2}=1.19
\end{equation}

Therefore:
\begin{equation}
<\epsilon_{\mu}> \sim exp^{-T^{2}/20}
\end{equation}

On the other hand, taking into account the relationship obtained for
the ratio between both efficiencies, $\epsilon_{\mu}$ and
$\epsilon_{H}$, this latter one may also be expressed by a similar
function : $\epsilon_{H} \propto exp^{-T^{2}/B}$ where B would be 8,
following our hypothesis from Section 2.

The result that efficiencies are related to galaxy morphological type
may be useful when a model for a given galaxy must be chosen and the
observational data are scarce. In that case, once the total mass
distribution has been selected from the rotation curve or the
luminosity for the galaxy, an initial choice can be made under the
assumption that the galaxy evolution is represented by the model of
efficiencies N with $N=T$. We warn, however, that given the large
dispersion of the data around the averaged efficiency values in
Fig.~\ref{emu1}, other values of N are also possible.

This finding would imply that the features of spiral galaxies depend
at least on two parameters: total mass and morphological type, as some
studies about the characteristics of galaxies, as the seminal work by
\cite{rob94}, show \citep[see also][among those more related with the
chemical evolution of galaxies]{vil92,zar94,gar97}.

It is also well known that the radial gradient, measured as dex $\rm
kpc^{-1}$, observed in spiral galaxies depends on morphological type:
late-type galaxies show steeper radial distributions of oxygen
abundances than earlier ones, which show in some cases almost no
gradient \citep{oey93,dut99}.  At the same time, a correlation seems
to exist between oxygen abundance and total mass surface density
which, in turn, is related to morphological type. This correlation is
stronger when the total mass of the galaxy, including the bulge,
instead of the mass of the exponential disc alone, is used
\citep{ryd95}. It is evident from these works that a linear sequence
with the mass can not be obtained, and therefore our models, being a
bi-parametric grid, would be adequate to find the best model for each
galaxy, although the existing relation between the morphological type
of galaxies and their total mass does difficult to discriminate which
of these features is the origin of the different evolution of galaxies.

From the chemical evolution point of view, a recent work (Moll\'{a} \&
M\'{a}rquez, in preparation), where the multiphase evolution model has
been applied to a large sample (67) of spiral galaxies, seems to
demonstrate that the total mass, through its influence on the collapse
time scale, has a larger effect on the evolution of a galaxy and its
final radial gradient of abundances than the selected values of the
named efficiencies. Thus, it seems as if galaxies would evolve mostly
due to their total mass with some dispersion around the mean trend
depending on hydrodynamical and environmental characteristics, which
are taken into account in some way by the efficiencies values. The
morphological type would then be the consequence of the different
conditions of the intergalactic medium out of which the galaxy formed.

\section{Summary and conclusions}

We have calculated the chemical evolution of a wide set of theoretical
galaxies characterised by their total mass, through the collapse time
scale, and their efficiencies to form molecular clouds and massive
stars from cloud-to-cloud collisions.

With the selection of parameters and inputs described above, we have
ran a total of 440 models, with 44 different rotation curves
--implying 44 values of total mass, characteristic collapse time-scale
and disc radius-- and 10 sets of efficiencies for each one of them,
implying 10 evolutionary rates for the star formation and gas
consumption in the disc.

For each model we have obtained the time evolution of the halo and the
disc, and therefore the corresponding radial distributions for the
relevant quantities (masses, abundances, star formation rate, etc...).
The star formation history for each radial region (halo and disc,
separately) and within each one, the mass in each phase of matter:
diffuse gas, molecular gas, low-mass stars, massive stars and
remnants, has been followed.  Besides that, we have obtained the
abundances of 15 elements: H, D, He3, He4, C12, C13, N14, O16, Ne, Mg,
Si, Ca, S, Fe and neutron-rich nuclei, in all radial regions for both
halo and disc.

The results of our work can be summarised as follows:

\begin{enumerate}
\item The atomic gas shows a maximum in its radial distribution for
all galaxies. This maximum is nearer to the galaxy centre in the low
mass or less evolved galaxies than in the more evolved or massive
galaxies, for which the maximum is along the disc and moving toward
the outer zones. This behavior produces, in some cases, a central {\sl
hole} in the distribution of the diffuse gas. The diffuse gas radial
distribution results to be a very strong constraint for selecting the
best model out of the 10 computed ones with different efficiencies,
corresponding to the total mass of a given spiral galaxy.

\item The oxygen abundance reaches a maximum level, as a consequence
of a saturation effect which occurs earlier for the massive and more
evolved galaxies. The less evolved galaxies do not reach this
saturation level, except in the central region, and therefore show a
steep radial gradient in their oxygen abundance. The less massive and
less evolved galaxies have not yet developed a radial gradient and
show flat radial distributions. This simulates an on-off effect: for
$N=7$ a radial gradient appears if $\lambda > 0.15$ while at $N=8 $ it
only appears for $\lambda \sim 1.50$.  This behavior is in agreement
with observations and solves the apparent inconsistency shown by
trends showing steep gradients for late type galaxies and flatter ones
for the earliest ones, while, at the same time, most irregulars show
no gradient at all and very uniform abundances.

\item The model calculated star formation rates reproduce the observed
trend between the surface density of this quantity and the total gas
density including, not only the massive normal galaxies data, but also
the low surface brightness ones.

\end{enumerate}

Actually, to our knowledge, there are not other chemical evolution
models which compare predicted with observed radial distributions of
diffuse and molecular gas, star formation rate and abundances for disc
galaxies other than MWG.  In this sense our models should be
considered as an improvement over the standard ones.

A study of the possible correlations among galaxy properties and
different features obtained from the complete set of results is now
possible, since a statistically significant number of theoretical
models is available.  This requires a deeper analysis which is out of
the scope of this paper and will be done in the next future.

\section*{Acknowledgments}

We thank an anonymous referee for many useful comments and suggestions
that have greatly improved this paper.  This work has been partially
funded by the Spanish Ministerio de Ciencia y Tecnolog\'{\i}a through
project AYA-2000-093. This work has made use of the Nasa Astrophysics
Data System, and the NASA/IPAC Extragalactic Database (NED), which is
operated by the Jet Propulsion Laboratory, Caltech, under contract
with the National Aeronautics and Space Administration.

\newpage
\setcounter{table}{4}  
\footnotesize
\begin{table*}
\begin{minipage}{126mm}
\caption{Model Results: Elemental Abundances.}
\label{abundances}
\begin{tabular}{rrrrrrrrrrrrrrrr}
\hline
\noalign{\smallskip}
Time  & R  & H & D & $^{3}$He & $^{4}$He & $^{12}$C & 
      $^{13}$C & N &  O       & Ne       &  Mg      & 
      Si       & S & Ca       & Fe \\
\hline
\noalign{\smallskip} 
  ... & ... & ...  & ...  & ...  & ...  & ...  & ...  & ...  & ...  & ...  & ...  & ...  & ...  & ...  & ...  \\
   0.1 &14. &0.770 & 0.70E--04 & 0.10E--04 &0.230 & 0.11E--06 & 0.21E--09 & 0.13E--07 & 0.92E--06 & 0.23E--06 & 0.32E--07 & 0.55E--07 & 0.26E--07 & 0.37E--08 & 0.42E--07 \\
   0.1 &12. &0.770 & 0.70E--04 & 0.10E--04 &0.230 & 0.12E--06 & 0.22E--09 & 0.13E--07 & 0.95E--06 & 0.23E--06 & 0.34E--07 & 0.57E--07 & 0.27E--07 & 0.38E--08 & 0.44E--07 \\
   0.1 &10. &0.770 & 0.70E--04 & 0.10E--04 &0.230 & 0.13E--06 & 0.23E--09 & 0.14E--07 & 0.10E--05 & 0.25E--06 & 0.35E--07 & 0.60E--07 & 0.28E--07 & 0.40E--08 & 0.46E--07 \\
   0.1 & 8. &0.770 & 0.70E--04 & 0.10E--04 &0.230 & 0.14E--06 & 0.25E--09 & 0.15E--07 & 0.11E--05 & 0.27E--06 & 0.38E--07 & 0.65E--07 & 0.31E--07 & 0.43E--08 & 0.50E--07 \\
   0.1 & 6. &0.770 & 0.70E--04 & 0.10E--04 &0.230 & 0.15E--06 & 0.28E--09 & 0.17E--07 & 0.12E--05 & 0.30E--06 & 0.43E--07 & 0.73E--07 & 0.35E--07 & 0.49E--08 & 0.56E--07 \\
   0.1 & 4. &0.770 & 0.70E--04 & 0.10E--04 &0.230 & 0.18E--06 & 0.32E--09 & 0.19E--07 & 0.14E--05 & 0.35E--06 & 0.50E--07 & 0.85E--07 & 0.40E--07 & 0.57E--08 & 0.65E--07 \\
   0.1 & 2. &0.770 & 0.70E--04 & 0.10E--04 &0.230 & 0.45E--06 & 0.46E--09 & 0.28E--07 & 0.38E--05 & 0.95E--06 & 0.13E--06 & 0.22E--06 & 0.10E--06 & 0.14E--07 & 0.15E--06 \\
  ... & ... & ...  & ...  & ...  & ...  & ...  & ...  & ...  & ...  & ...  & ...  & ...  & ...  & ...  & ...  \\
  13.2 &14. &0.769 & 0.69E--04 & 0.12E--04 &0.231 & 0.59E--04 & 0.12E--06 & 0.68E--05 & 0.15E--03 & 0.34E--04 & 0.52E--05 & 0.10E--04 & 0.50E--05 & 0.71E--06 & 0.24E--04 \\
  13.2 &12. &0.768 & 0.69E--04 & 0.13E--04 &0.231 & 0.10E--03 & 0.21E--06 & 0.96E--05 & 0.27E--03 & 0.57E--04 & 0.84E--05 & 0.20E--04 & 0.96E--05 & 0.13E--05 & 0.42E--04 \\
  13.2 &10. &0.764 & 0.65E--04 & 0.21E--04 &0.234 & 0.43E--03 & 0.98E--06 & 0.38E--04 & 0.11E--02 & 0.21E--03 & 0.30E--04 & 0.86E--04 & 0.40E--04 & 0.55E--05 & 0.17E--03 \\
  13.2 & 8. &0.754 & 0.58E--04 & 0.44E--04 &0.240 & 0.11E--02 & 0.44E--05 & 0.16E--03 & 0.28E--02 & 0.54E--03 & 0.76E--04 & 0.22E--03 & 0.10E--03 & 0.14E--04 & 0.44E--03 \\
  13.2 & 6. &0.746 & 0.51E--04 & 0.73E--04 &0.246 & 0.16E--02 & 0.11E--04 & 0.37E--03 & 0.40E--02 & 0.78E--03 & 0.11E--03 & 0.33E--03 & 0.16E--03 & 0.21E--04 & 0.67E--03 \\
  13.2 & 4. &0.736 & 0.43E--04 & 0.13E--03 &0.253 & 0.21E--02 & 0.22E--04 & 0.64E--03 & 0.49E--02 & 0.98E--03 & 0.14E--03 & 0.44E--03 & 0.21E--03 & 0.28E--04 & 0.93E--03 \\
  13.2 & 2. &0.693 & 0.63E--05 & 0.43E--03 &0.287 & 0.37E--02 & 0.71E--04 & 0.18E--02 & 0.81E--02 & 0.18E--02 & 0.26E--03 & 0.83E--03 & 0.40E--03 & 0.56E--04 & 0.20E--02 \\
\hline
\noalign{\smallskip}
\end{tabular}
\end{minipage}
\end{table*}
\normalsize
\label{lastpage}

\end{document}